\documentclass[aps,prd,twocolumn,nofootinbib,twoside,floatfix,reprint]{revtex4-1}

\usepackage{aas_macros}
\usepackage{natbib}
\usepackage{amsmath}
\usepackage{amssymb}
\usepackage{graphicx}
\usepackage{microtype}
\usepackage{color}

\newcommand{\p}{\ensuremath{\partial}}
\newcommand{\del}{\ensuremath{\delta}}
\newcommand{\Del}{\ensuremath{\Delta}}

\newcommand{\gam}{\ensuremath{\gamma}}

\newcommand{\sig}{\ensuremath{\sigma}}
\newcommand{\Om}{\ensuremath{\Omega}}

\newcommand{\delc}{\ensuremath{\delta_{\rm c}}}

\def\Mpc{\, h^{-1}{\rm Mpc}}

\def\Ms{h^{-1}M_{\odot}}

\newcommand{\avg}[1]{\ensuremath{\left\langle \,#1\, \right\rangle}}

\newcommand{\der}{\ensuremath{{\rm d}}}

\newcommand{\erf}[1]{\ensuremath{{\rm erf}\left(#1\right)}}

\newcommand{\eqn}[1]{equation~\eqref{#1}}
\newcommand{\eqns}[1]{equations~\eqref{#1}}
\newcommand{\fig}[1]{Figure~\ref{#1}}

\newcommand{\ph}[1]{\phantom{#1}}

\newcommand{\be}{\begin{equation}}
\newcommand{\ee}{\end{equation}}
\newcommand{\Cal}[1]{\ensuremath{\mathcal{#1}}}

\begin{document}
\title{Cosmology with Galaxy Clusters: Systematic Effects in the Halo Mass Function}
\author{Aseem Paranjape}
\email{aseemp@phys.ethz.ch}
\affiliation{Institute for Astronomy, Department of Physics, ETH Z\"urich, 
  Wolfgang-Pauli-Strasse 27, CH 8093 Z\"urich, Switzerland\\}
\begin{abstract}
\noindent
We investigate potential systematic effects in constraining the amplitude of primordial fluctuations $\sigma_8$ arising from the choice of halo mass function in the likelihood analysis of current and upcoming galaxy cluster surveys. 
We study the widely used $N$-body simulation fit of Tinker et al. (T08) and, as an alternative, the recently proposed analytical model of Excursion Set Peaks (ESP). 
We first assess the relative bias between these prescriptions when constraining $\sigma_8$ by sampling the ESP mass function to generate mock catalogs and using the T08 fit to analyse them, for various choices of survey selection threshold, mass definition and statistical priors.
To assess the level of absolute bias in each prescription, we then repeat the analysis on dark matter halo catalogs in $N$-body simulations designed to mimic the mass distribution in the current data release of Planck SZ clusters. 
This $N$-body analysis shows that using the T08 fit \emph{without} accounting for the scatter introduced when converting between mass definitions (alternatively, the scatter induced by errors on the parameters of the fit) can systematically  \emph{over-estimate} the value of $\sigma_8$ by as much as $2\sigma$ for current data, while analyses that account for this scatter should be close to unbiased in $\sigma_8$. With an increased number of objects as expected in upcoming data releases, regardless of accounting for scatter, the T08 fit could over-estimate the value of $\sigma_8$ by $\sim1.5\sigma$. 
The ESP mass function leads to systematically more biased but comparable results.
A strength of the ESP model is its natural prediction 
of a weak non-universality in the mass function which closely tracks the one measured in simulations and described by the T08 fit. We suggest that it might now be prudent to build new unbiased ESP-based fitting functions for use with the larger datasets of the near future. 
\end{abstract}

\maketitle
\section{Introduction}
\label{sec:intro}
\noindent
Cosmology is now a precision science. The wealth of cosmological data from measurements of the Cosmic Microwave Background (CMB), Large Scale Structure and related probes is well described by the simple $6$-parameter Lambda-Cold dark matter ($\Lambda$CDM) model, whose parameters are now known with unprecedentedly small errors. The last decade in particular has witnessed a ten-fold increase in precision in recovering the values of these parameters \cite{jaffe01,planck13-XVI-cosmoparams}. Cosmological analyses have reached the stage where the error budget on parameter constraints is starting to be dominated by systematic rather than statistical uncertainties. Understanding these systematic effects -- in both data analysis as well as theoretical modeling -- is a pressing challenge, particularly in light of assessing the importance of tensions when constraining a given parameter from different data sets and complementary probes.

We focus here on cosmological constraints from the abundance of galaxy clusters (see \cite{borgani08,aem11} for reviews). The sensitivity of cluster number counts to parameters such as $\sig_8$ (the strength of the primordial density fluctuations) and $\Om_{\rm m}$ (the fractional budget of non-relativistic matter) means that these remain a competitive probe even today \cite{hhm01,bw03,mb06,sahlen+09,chf09,fcmc11,weinberg+13}. 
Recent results from the Planck Collaboration \cite{planck13-XX-SZcosmo} suggest that there is a $2$-$3\sigma$ tension between the value of $\sig_8$ recovered from measurements of the CMB and from galaxy cluster counts determined using the Sunyaev-Zel'dovich (SZ) effect. It has been suggested that this tension could arise due to systematic choices in the CMB data analysis pipeline \cite{sfh13}, or due to mis-calibration of the mass-observable relation \cite{planck13-XX-SZcosmo,cbm14,vdL+14}, or even through more non-standard effects such as those due to massive neutrinos \cite{planck13-XX-SZcosmo,hh13,bm13} (although see \cite{castorina+13}).

In this paper we investigate another potential source of systematic biases, namely, the halo mass function. The complexity of the nonlinear gravitational effects that lead to the formation of gravitationally bound, virialised `halos' has meant that, despite considerable analytical progress over the last several years, the gold standard for estimating the halo mass function continues to be measurements in numerical simulations. In addition to accounting for this complexity, simulations also allow for calibrations of the mass function for the various choices of mass definition that are suited to the specific observational probe (such as SZ flux/X-ray luminosity/optical richness) rather than being restricted to theoretical approximations and assumptions such as spherical or ellipsoidal collapse (see \cite{bk09} for a review).

However, the nature of parameter recovery through likelihood maximisation or Bayesian techniques  
means that it is crucial to use analytical approximations that accurately capture the effect of cosmology on the mass function. Since it is unfeasible to run an $N$-body simulation for every combination of parameter values,
the standard compromise has been the use of analytical fits to the results of simulations \cite{st99,jenkins+01,Tinker08,watson+13} (although, in principle, it should be possible to directly interpolate between simulations along the lines of \cite{heitmann+14,kwan+13}). As we emphasize below, these fits are routinely used in analyses of cluster abundances \emph{without} accounting for the error covariance matrices of the fit parameters \cite{vanderlinde+10,benson+13,hasselfield+13,reichardt+13,planck13-XX-SZcosmo}, and this opens the door to potential systematic biases \cite{ce10,bhattacharya+11,pmw13}. 

In the following we will set up a pipeline for analysing mock cluster catalogs, including various choices of survey selection threshold, mass-observable relation and priors on cosmological parameters, with a focus on the effect of the halo mass function model. Our catalogs will be based on both Monte Carlo sampling of analytical mass functions as well as halos identified directly in $N$-body simulations of CDM, and will allow us to explore the interplay between the nonlinear systematics inherent in the chosen mass function model and the other ingredients mentioned above.
Although we do not explicitly model baryonic effects (these are expected to systematically alter the mass function at the $10$-$20\%$ level; see, e.g., \cite{stanek+10,bp13,martizzi+14,cbm14,velliscig+14}), our examples below will include biased mass-observable relations that show similar features.

The paper is organised as follows. In Section~\ref{sec:analytical} we discuss the main analytical approximations used in typical cluster analyses, namely, the cluster likelihood and the halo mass function. We will focus on two prescriptions for the latter, namely the $N$-body fits of \citet{Tinker08} and the theoretical Excursion Set Peaks (ESP) prescription of \cite{psd13}. In Section~\ref{sec:montecarlo} we perform an in-depth statistical comparison of the $N$-body fits and the ESP mass function by using the former to analyse Monte Carlo mock catalogs generated by sampling the latter. In Section~\ref{sec:Nbody} we repeat the analysis using both these prescriptions to analyse catalogs built from halos identified in $N$-body simulations of CDM that were designed to mimic the mass distribution in the current data release of Planck SZ clusters. We conclude in Section~\ref{sec:conclude}. Appendix~\ref{app:masses} gives various technical details regarding mass calibration issues while Appendix~\ref{app:lightcones} describes our procedure for generating lightcones from the $N$-body halos.

We assume a flat $\Lambda$CDM cosmology with Gaussian initial conditions. Unless stated otherwise, for our fiducial cosmology we set the fraction of total matter $\Om_{\rm m}=0.315$, the baryonic fraction $\Om_{\rm b}=0.0487$, the Hubble constant $H_0=100h\,{\rm km/s/Mpc}$ with $h=0.673$, the scalar spectral index $n_s=0.96$ and the linearly extrapolated r.m.s. of matter fluctuations in spheres of radius $8\Mpc$, $\sig_8=0.83$, which are compatible with the analysis of Planck CMB data \cite{planck13-XVI-cosmoparams}. We use the transfer function prescription of \citet{eh98} for all our calculations. We denote the natural logarithm of $x$ by $\ln(x)$ and the base-10 logarithm by $\log(x)$.

\section{Analytical approximations}   
\label{sec:analytical}
\noindent
The primary ingredients in the statistical modeling of cluster number counts are the likelihood as a function of redshift and mass-proxy (including the effects of the survey selection threshold), and the halo mass function. 
We discuss each of these below.

\subsection{Likelihood for Cluster Cosmology}
\label{sec:analytical:sub:likelihood}
\noindent
The likelihood for cluster abundances is built in several steps. Given the mass function $\der n/\der\ln m(m,z)$, i.e. the comoving number density of halos with logarithmic masses in the range $(\ln m,\ln m+\der\ln m)$ at redshift $z$, the expected number of halos in this mass range and in the redshift range $(z,z+\der z)$ is
\be
f_{\rm sky}\, \der z\, \der\ln m\, \frac{\der V}{\der z}\frac{\der n}{\der\ln m}
\notag
\ee
where $f_{\rm sky}$ is the sky fraction covered by the survey\footnote{For simplicity we ignore variations in the survey depth as a function of angle in the sky.} 
and $\der V/\der z = 4\pi H(z)^{-1}(\int_0^z\der z'/H(z'))^2$ is the cosmology-dependent volume function with $H(z)$ the Hubble parameter in units of $h/{\rm Mpc}$. Below we will consider a Planck-like survey for which we set $f_{\rm sky}=0.48$ consistent with the current release of Planck SZ clusters \cite{planck13-XXIX-SZcatalog}, and a South Pole Telescope (SPT)-like survey for which we set $f_{\rm sky}=0.06$ consistent with the results expected from the full analysis of SPT data (note that the latest data release covers the first $720\,$sq. deg., or $f_{\rm sky}=0.0173$ \cite{reichardt+13}).

The next step is to connect the halo mass $m$ to the observable $Y$; this could be the Sunyaev-Zel'dovich flux $Y_{\rm SZ}$ for SZ-detected clusters \cite{planck13-XX-SZcosmo}, the X-ray luminosity $L_X$ \cite{rykoff+08} or the product $Y_X$ of X-ray temperature and gas mass \cite{kvn06} for X-ray observations, or the richness of optically detected clusters \cite{gladders+07,koester+07}. This is done by modeling a stochastic relation $p(Y|m)$ between $Y$ and $m$, typically assumed to be a Lognormal in $Y$ with mean scaling relation $\avg{\ln Y|\ln m} = \overline{\ln Y}(\ln m)$ and scatter $\sig_{(\ln Y|\ln m)}$ calibrated to simulations.
A particularly thorny issue, which has received much attention \cite{stanek+10,bp13,cbm14,vdL+14}, is the need to calibrate possible biases in the scaling relation $\overline{\ln Y}(\ln m)$.
A typical method for dealing with such a bias is to introduce an additive constant in the relation $\overline{\ln Y}(\ln m)$ which could then be fit simultaneously with the cosmological parameters \cite{planck13-XX-SZcosmo}. 

In this work, we are interested in theoretical systematic effects that could enter through inaccuracies in modeling the mass function $\der n/\der\ln m$, and \emph{not} with any systematic effects that enter through the step that relates $m$ to $Y$. To this end we replace $Y$ with $m_{\rm ob}$ (an ``observed mass'' or mass proxy), and consider various choices for $m_{\rm ob}$ such as $m_{\rm 500c}$ or $m_{\rm 200b}$ (defined in Section~\ref{sec:analytical:sub:Nbodyfits}), the distributions of which are reliably accessible in numerical simulations. We will nevertheless use the statistical language mentioned above in order to, at least formally, connect with data analyses that do model the $m$-$Y$ relation. E.g., we will study the effects of biases similar to those mentioned above by modeling the stochastic relations between different mass definitions.

Finally, the survey completeness function $\chi(Y,z)$ gives the probability that a cluster with observable value $Y$ at redshift $z$ will be seen in a survey, given that the cluster exists\footnote{We will assume that there are no false positive detections, although we note that impurities in the sample can also affect the mass distribution near the selection threshold.}. 
One simplification we will use is to model the function $\chi(m_{\rm ob},z)$ as being unity for $m_{\rm ob}$ larger than a suitably defined survey selection threshold $M_{\rm ob,lim}(z)$ and zero otherwise. We will then use this \emph{fixed} threshold evaluated in the fiducial cosmology to both define our numerical catalogs as well as analyse them. This allows for a straightforward comparison of the analytical and numerical mass functions. In principle the analysis could be made more realistic by allowing for smoothly varying functions $\chi(m_{\rm ob},z)$; we will not explore this here.

We motivate the choice of threshold $M_{\rm ob,lim}(z)$ by approximating the observed mass distributions in the Planck and SPT surveys. We find that the following functional form provides a reasonable description\footnote{Although the shape of the selection threshold \eqref{Moblim-Planck} can be motivated using the scaling relation in equation~(7) of \cite{planck13-XX-SZcosmo} evaluated at a fixed value of $\chi$ and noise $\sig_{Y_{500}}$, a proper derivation would actually involve self-consistently solving equations~(7) and~(8) of \cite{planck13-XX-SZcosmo} (which relate the observable $Y_{500}$ and angular aperture $\theta_{500}$, respectively, to the mass $m_{\rm 500c}$) along with a relation $\sig_{Y_{500}}(\theta_{500})$ describing the noise as a function of aperture. Since the latter is not provided in \cite{planck13-XX-SZcosmo}, we resort to \eqn{Moblim-Planck} which approximately matches the green curve for the `shallow zone' in Figure~3 of \cite{planck13-XX-SZcosmo}.}
of the Planck selection threshold for the mass $m_{\rm 500c}$:
\begin{align}
M_{\rm ob,lim}^{\rm (Planck)}(z) &= 9.14\times10^{13}\Ms\notag\\
&\ph{10^{15}}
\times E(z)^{-\beta/\alpha}\times\left(\frac{D_A(z)}{100\Mpc}\right)^{2/\alpha}\,,
\label{Moblim-Planck}
\end{align}
where $D_A(z)$ is the angular diameter distance to redshift $z$ and $E(z)\equiv H(z)/H_0$ is the normalised Hubble parameter in our fiducial cosmology, and we set $\beta=0.66$, $\alpha=1.79$ (Table~1 of \cite{planck13-XX-SZcosmo}).

With this choice we find that the expected number of clusters (using the ESP mass function described below) is $\sim184$ for our fiducial value of $f_{\rm sky}=0.48$ for such a survey, which is reassuringly close to the actual number of clusters analysed by the Planck Collaboration which is $189$. The steepness of the mass function means, however, that the systematic effects we study below could in principle depend sensitively on the specific choice of selection threshold. To ensure that our final conclusions are robust to uncertainties in this choice, we will later also quote results for a slightly modified version of \eqref{Moblim-Planck}.

For an SPT-like survey the limiting mass is approximately independent of redshift above $z\gtrsim0.3$ \cite{reichardt+13}. When discussing results for such a survey we will set 
\be
M^{\rm (SPT)}_{\rm ob,lim} = 3.2\times10^{14}\Ms\,;\quad z\geq0.3\,,
\label{Moblim-SPT}
\ee
for which the expected number of clusters using the ESP mass function is $\sim470$ for our fiducial value of $f_{\rm sky}=0.06$ (i.e., $2500\,$sq. deg). For $720\,$sq. deg. this gives a fiducial count of $\sim135$ clusters, close to the number actually observed by SPT which is $158$. 

Note that our choice of treating $m_{\rm ob}$ as fundamental -- rather than additionally modeling the relation to an observable $Y$ -- allows us to ignore the cosmology dependence of $M_{\rm ob,lim}(z)$ when performing the likelihood analysis, and also means that our analysis is not affected by Malmquist bias when using the selection thresholds~\eqref{Moblim-Planck} and~\eqref{Moblim-SPT} since we will directly use these thresholds when defining our numerical catalogs.

Putting things together, the expected number of clusters in the $i^{\rm th}$ mass bin and $j^{\rm th}$ redshift bin in the $(m_{\rm ob},z)$ plane with boundaries $m_{{\rm ob},i}^{-} < m_{\rm ob} \leq m_{{\rm ob},i}^{+}$ and $z_j^{-} < z \leq z_j^{+}$, is given by
\begin{align}
\mu_{ij} &=  \mu(m_{{\rm ob},i},z_j)\notag\\
&= f_{\rm sky}\int_{z\,{\rm bin\,}j}\der z\frac{\der V}{\der z}\int\der\ln m \frac{\der n}{\der\ln m}\notag\\
&\ph{f_{\rm sky}}\times
\int_{{\rm mass\,bin\,}i}\der\ln m_{\rm ob}\,p(\ln m_{\rm ob}|\ln m)\chi(m_{\rm ob},z)\notag\\
&= f_{\rm sky}\int_{z_j^{-}}^{z_j^{+}}\der z\frac{\der V}{\der z}\int\der\ln m \frac{\der n}{\der\ln m}\notag\\
&\ph{f\int}\times 
\frac12\bigg[\erf{\frac{\avg{\ln m_{\rm ob}|\ln m} - \ln m_{{\rm ob},i}^{-}}{\sqrt{2}\sig_{\ln m_{\rm ob}}}}\notag\\
&\ph{f\int\frac12[]} 
 - \erf{\frac{\avg{\ln m_{\rm ob}|\ln m} - \ln m_{{\rm ob},i}^{+}}{\sqrt{2}\sig_{\ln m_{\rm ob}}}}\bigg]\,,
\label{mu-mobz}
\end{align}
where, in the last equality, we have peformed the integration of a Lognormal distribution in $m_{\rm ob}$ over the mass bin which is understood to be above the threshold mass. Hereafter, for convenience we will denote $\sig_{(\ln m_{\rm ob}|\ln m)}$ as simply $\sig_{\ln m_{\rm ob}}$. We discuss our choices for $\avg{\ln m_{\rm ob}|\ln m}$ and $\sig_{\ln m_{\rm ob}}$ later.

Finally, one assumes that the actual number of clusters $N_{ij}$ observed in the bin is a Poisson realisation with mean $\mu_{ij}$, and that individual bins are uncorrelated with each other, which is a good approximation for large enough surveys and redshift bins \cite{hk03,hc06}. This gives the likelihood
\be
\Cal{L} = \prod_{i,j} \frac{\mu_{ij}^{N_{ij}}}{N_{ij}!}{\rm e}^{-\mu_{ij}}\,.
\label{likelihood}
\ee
We remark in passing that this is not equivalent to first summing over all mass bins above the limiting mass for any redshift $z_j$ and then writing $\tilde{\Cal{L}}=\prod_j{\rm e}^{-\mu_j}\mu_j^{N_j}/N_j!$ with $\mu_j$ given by integrating $\mu_{ij}$ over masses $m_{\rm ob} > M_{\rm ob,lim}(z_j)$. For the same values of the cosmological parameters, the likelihood $\tilde{\Cal{L}}$ allows for many more combinations of mass distributions at fixed redshift than does $\Cal{L}$, and it is easy to show that $\tilde{\Cal{L}} > \Cal{L}$ always. In general this would mean that $\Cal{L}$ is more constraining than $\tilde{\Cal{L}}$; however, the exact influence of this choice on the significance of biases induced by nonlinear systematics is difficult to judge. 
In this work we will use only equation~\eqref{likelihood}, since this uses the maximum available information from the survey. We note that the Planck Collaboration have chosen to work with $\tilde{\Cal{L}}$ instead, presumably because this is likely to be less sensitive to inaccuracies in modeling the relation $p(\ln m_{\rm ob}|\ln m)$ which controls the leakage of objects across bins.

\subsection{Halo mass function: Original excursion set approach}
\label{sec:analytical:sub:excsets}
\noindent
A key requirement for analysing the likelihood described above is a sufficiently accurate analytical prescription for computing the halo mass function $\der n/\der\ln m$. All current models for the mass function, including fits to $N$-body simulations, are built upon the so-called excursion set approach which we briefly describe here \cite{ps74,e83,ph90,bcek91,lc93,mw96,bm96,smt01}.

This approach makes the ansatz that `sufficiently dense' patches in the initial conditions of the Universe can be mapped to virialised halos at the epoch of interest. The criterion for being sufficiently dense follows from making approximations to the nonlinear gravitational dynamics such as spherical \cite{gg72,ps74} or ellipsoidal collapse \cite{bm96,dpg98,Monaco99,smt01}. The halo mass function is then written as
\be
\frac{\der n}{\der\ln m} = \frac{\bar\rho(0)}{m}\,\nu f(\nu) \left|\frac{\der\ln\nu}{\der\ln m}\right|\,,
\label{mf-basic}
\ee
with $\bar\rho(0)$ the mean matter density of the Universe at $z=0$ and where $f(\nu)$ is an output of the excursion set calculation and gives the mass fraction in collapsed objects in terms of the scaling variable $\nu$ defined as
\be
\nu(m,z)\equiv\frac{\delc(z)}{\sig_0(m)}\frac{D(0)}{D(z)}\,.
\label{nu-def}
\ee
Here $\delc(z)$ is the critical collapse threshold (for the linearly extrapolated density contrast) in the spherical collapse model\footnote{The value of $\delc(z)$ in a flat $\Lambda$CDM universe is weakly dependent on redshift and cosmology, in contrast to that in an Einstein-deSitter background (see, e.g., \cite{ecf96}), and can be approximated by $\delc(z) = \del_{\rm c,EdS}(1-0.0123\log_{10}(1+x^3))$, where $x\equiv(\Omega_{\rm m}^{-1}-1)^{1/3}/(1+z)$ and $\del_{\rm c,EdS}=1.686$ \citep{Henry2000}. For example, requiring collapse at present epoch for our fiducial cosmology gives $\delc(z=0)=1.675$.}, $D(z)$ is the linear theory growth factor and $\sig_0^2(m)=\avg{\del_R^2}$ is the variance of the density contrast smoothed on comoving scale $R$ such that $m=4\pi R^3\bar\rho(0)/3$,
\be
\sig_0^2(m) = \int\der\ln k\,\Del^2(k)\,W(kR)^2\,,
\label{sig0-def}
\ee
where $\Del^2(k)\equiv k^3P(k)/(2\pi^2)$ is the dimensionless matter power spectrum, linearly extrapolated to $z=0$, and $W(kR)$ is the Fourier transform of the real-space spherical TopHat filter: $W(x)=(3/x^3)(\sin x-x\cos x)$.

The classic calculation of the mass fraction \cite{ps74,bcek91} identifies halos of mass $m$ with regions in the initial density that are dense enough to collapse when smoothed on scale $R\propto m^{1/3}$ but not on any larger scale, and gives the well-known Press-Schechter \cite{ps74} result
\be
\nu f_{\rm PS}(\nu) = \sqrt{2/\pi}\,\nu\,{\rm e}^{-\nu^2/2}\,,
\label{vfv-PS}
\ee
which is the distribution of scales at which random walks in the linearly extrapolated density contrast, as a function of decreasing smoothing scale or increasing $\sig_0(m)$, first cross the `barrier' $\delc(z)D(0)/D(z)$ \cite{bcek91}.

\subsection{Halo mass function: $N$-body fits}
\label{sec:analytical:sub:Nbodyfits}
\noindent
Traditionally, the original excursion set results \cite{ps74,bcek91} and their extensions to mass-dependent collapse thresholds \cite{smt01} have been used as templates to fit the mass functions measured in $N$-body simulations after introducing some free parameters. E.g., the \citet[][hereafter, ST99]{st99} fitting function is
\be
\nu f_{\rm ST}(\nu) = \tilde A\,\sqrt{2q/\pi}\,\nu\,{\rm e}^{-q\nu^2/2}\left[1+(q\nu^2)^{-p}\right]\,,
\label{vfv-ST}
\ee
where $\tilde A=(1+\Gamma(1/2-p)2^{-p}/\sqrt{\pi})^{-1}$ ensures normalisation and $\Gamma(x)$ is the Euler gamma function. ST99 reported that the parameter values $q=0.707$ and $p=0.3$ gave a good fit to the mass function of halos identified using a Spherical Overdensity (SO) criterion \cite{lc94} (see below) in the GIF simulations \cite{kcdw99}. Since that early work, there have been a number of calibrations of the SO as well as Friends-of-Friends (FoF) \cite{defw85} mass functions, spanning larger ranges in mass and redshift \cite{jenkins+01,warren+06,lukic+07,Tinker08,pph10,mice10,bhattacharya+11,watson+13}.

For cluster surveys it is useful to have calibrated mass functions for SO halos with masses determined by growing spheres around chosen locations (e.g., density peaks) until the spherically averaged dark matter density falls below a specific threshold. E.g., the SZ surveys we dicuss in this work typically use the definition $m_{\rm 500c}$ which is the mass enclosed in a sphere of radius $R_{\rm 500c}$ at which the enclosed dark matter density falls to $500$ times the critical density $\rho_{\rm c}(z)=3H^2(z)/(8\pi G)$ of the Universe. Another popular definition replaces the critical density with the background density $\bar\rho(z)=\Omega_{\rm m}(z)\rho_{\rm c}(z)=\bar\rho(0)(1+z)^3$, resulting in masses $m_{\Delta{\rm b}}$ in spheres of radius $R_{\Delta{\rm b}}$ with, say, $\Delta=200$ (which we study below). Different SO mass definitions can be related to one another given the form of the halo density profile \cite{White01,hk03}; we discuss the specific example of relating $m_{\rm 200b}$ with $m_{\rm 500c}$ below and in Appendix~\ref{app:masscal}.

\citet[][hereafter, T08]{Tinker08} calibrated the functional form\footnote{The notation in equations~(2) and~(3) of T08 is different from ours; their $f(\sig)$ corresponds to what we call $\nu f_{\rm T08}(\nu)$, and their $\sig(m,z)$ corresponds to our $\sig_0(m)D(z)/D(0)=\delc/\nu$ (see equation~\ref{nu-def}).}
\be
\nu f_{\rm T08}(\nu) = A\,{\rm e}^{-c\nu^2/\delc^2}\left[1+(b\nu/\delc)^a\right]
\label{vfv-Tinker}
\ee
to SO halos with $m_{\Delta{\rm b}}$ masses identified in a suite of CDM $N$-body simulations, for a range of values of $\Delta$, resulting in mass function fits that are accurate at the $\sim5\%$ level over the mass range $10^{11} < m / (\Ms) < 10^{15}$ and redshift range $0 \leq z \leq 1.25$. A key ingredient in their analysis was the fact that the parameters $A,a,b$ were allowed to vary with redshift, resulting in a mass function that is explicitly non-universal at the $10$-$20\%$ level; this non-universality was crucial in obtaining the accuracy quoted above. Specifically, T08 found the following redshift dependence
\begin{align}
&A(z)=A_0(1+z)^{-0.14}\,;\quad a(z)=a_0(1+z)^{-0.06}\notag\\
&b(z)=b_0(1+z)^{-\alpha}\,;\quad \log\alpha(\Delta)=-\left(\frac{0.75}{\log(\Delta/75)}\right)^{1.2}\,,
\label{T08-Aab}
\end{align}
with the $\Delta$-dependent values of $A_0,a_0,b_0,c$ given in Table~2 of T08. 
The T08 fits have been used by several groups for deriving cosmological constraints using cluster surveys \cite{rozo+10,vanderlinde+10,benson+13,hasselfield+13,reichardt+13,planck13-XX-SZcosmo}. 

\subsection{Halo mass function: Excursion Set Peaks}
\label{sec:analytical:sub:ESP}
\noindent
In parallel with the increasing accuracy of numerical fits, the analytical understanding of the halo mass function has also considerably improved over the last several years \cite{zh06,mr10,pls12,ms12,arsc13,fb13,ms14}. One particular set of calculations that we will discuss here is known as Excursion Set Peaks (ESP). This is based on ideas presented by \cite{ms12,ps12} (see also \cite{Bond1989}) and developed further by \citet[][hereafter, PSD13]{psd13}. This framework essentially identifies halos of mass $m$ with \emph{peaks} (rather than arbitrary patches) in the initial density that are dense enough to collapse when smoothed on scale $R\propto m^{1/3}$ but not on any larger scale. It therefore combines peaks theory (see \citet[][hereafter, BBKS]{bbks86} for an excellent exposition) and the excursion set approach \cite{aj90,bm96,h01,jsdm14-I}. 

The criterion for being sufficiently dense is modelled using a mass-dependent barrier
\be
B(\sig_0,z) = \delc(z)D(0)/D(z) + \beta\sig_0(m)\,,
\label{esp-barrier}
\ee
which is motivated by the ellipsoidal collapse model \cite{bm96,smt01}, and where $\beta$ is a stochastic variable with distribution $p(\beta)$ that we discuss below. The ESP calculation then gives the following prescription for the mass fraction (PSD13)
\be
\nu f_{\rm ESP}(\nu) = \int \der\beta\,p(\beta)\,\nu f_{\rm ESP}(\nu|\beta)\,,
\label{vfv-ESP}
\ee
where
\begin{align}
f_{\rm ESP}(\nu|\beta) &= (V/V_\ast) ({\rm e}^{-(\nu+\beta)^2/2}/\sqrt{2\pi})\notag\\
&\ph{V/V_\ast}\times
 \int_{\beta\gam}^\infty \der x\,\frac{x-\beta\gam}{\gamma\nu} F(x)\notag\\
&\ph{V/V_\ast\times\int}\times p_{\rm G}(x-\beta\gam - \gam\nu;1-\gam^2)\,.
\label{fESP-beta}
\end{align}
Here $V=m/\bar\rho(0)=4\pi R^3/3$ is the Lagrangian volume of the halo/peak-patch, $p_{\rm G}(y-\mu;\Sigma^2)$ is a Gaussian distribution in the variable $y$ with mean $\mu$ and variance $\Sigma^2$, $F(x)$ is the peak curvature function that describes the effects of averaging over peak shapes
\begin{align}
F(x)&=\frac12\left(x^3-3x\right)\left\{\erf{x\sqrt{\frac52}}+\erf{x\sqrt{\frac58}}
  \right\} \notag\\
&\ph{x^3-3x}
+ \sqrt{\frac2{5\pi}}\bigg[\left(\frac{31x^2}{4}+\frac85\right){\rm
    e}^{-5x^2/8} \notag\\
&\ph{\sqrt{x^3-3x+\frac2{5\pi}}[]}
+ \left(\frac{x^2}{2}-\frac85\right){\rm
    e}^{-5x^2/2}\bigg]\,,
\label{eqn-bbks-Fx}
\end{align}
(equations~A14--A19 in BBKS), and $V_\ast$ and $\gamma$ are spectral quantities defined as
\begin{align}
V_{\ast} = (6\pi)^{3/2} (\sig_{1{\rm G}}/\sig_{2{\rm G}})^3\quad;\quad \gam = \sig_{1{\rm m}}^2/(\sig_0\sig_{2{\rm G}})\,,
\label{gam-Vst}
\end{align}
using the spectral integrals
\begin{align}
\sig_{j{\rm G}}^2 &\equiv \int \der\ln k\, \Del^2(k)\,k^{2j}{\rm e}^{-k^2R^2_{\rm G}} ~~,~ j\geq1\,,\notag\\
\sig_{1{\rm m}}^2 &= \int \der\ln k\, \Del^2(k)\,k^2 {\rm e}^{-k^2R^2_{\rm G}/2}W(kR)\,.
\label{spectral}
\end{align}
The Gaussian smoothing scale $R_{\rm G}$ is fixed by requiring $\avg{\del_{\rm G}|\del_{\rm TH}}=\del_{\rm TH}$ (the subscripts denoting {\bf G}aussian and {\bf T}op{\bf H}at smoothing, respectively), i.e., $\avg{\del_{\rm G}\del_{\rm TH}}=\sig_0^2$, which in practice leads to $R_{\rm G}\approx0.46R$ with a slow variation (PSD13).

The distribution of $\beta$ can be determined by requiring consistency with \emph{measurements} of $B$ in the initial conditions of an $N$-body simulation. \citet{rktz09}, e.g., have performed such measurements for the same simulations analysed by T08, for the $m_{\rm 200b}$ mass definition. In practice this is done by tracing back the $N$-body particles corresponding to a specific halo identified at, say, $z=0$ to their locations in the \emph{initial conditions}, thus demarcating a `proto-halo' corresponding to this halo. The value of the initial density contrast (linearly extrapolated to $z=0$ in this case) averaged over this proto-halo gives an estimate of $B$ for this object. Doing this for all halos at $z=0$ leads to a numerical sample of the distribution of $B$ as a function of halo mass, which \citet{rktz09} showed was well approximated by a Lognormal with mean value proportional to $\sig_0$ (see also \cite{hp14}).

PSD13 showed that setting the distribution $p(\beta)$ in \eqn{vfv-ESP} to be Lognormal in $\beta$ with mean $\avg{\beta}=0.5$ (which is close to the prediction of the ellipsoidal collapse model) and variance ${\rm Var}(\beta)=0.25$ gives a self-consistent description of both the $m_{\rm 200b}$ mass function of T08 as well as the proto-halo density distribution of \citet{rktz09}, accurate at the $\sim10\%$ level. In addition, the same choice of $p(\beta)$ then leads to a prediction for (nonlinear) halo bias that is also accurate at the $\sim5$-$10\%$ level when compared with simulations \cite{psd13,psc+13}. 

\begin{figure}
\includegraphics[width=0.475\textwidth]{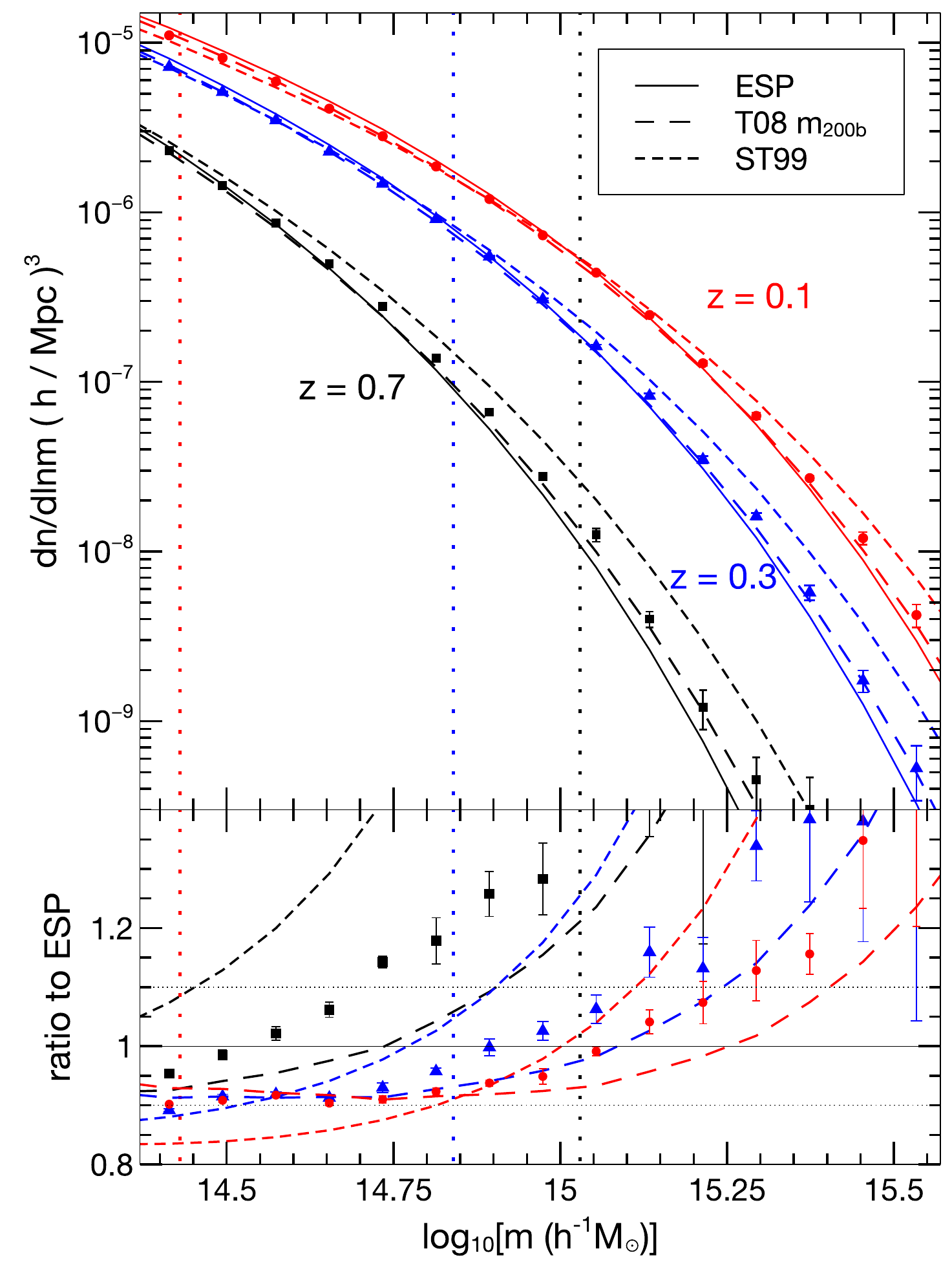}
\caption{Halo mass function at different redshifts. \emph{(Top panel):} Smooth curves show the mass function from the three prescriptions discussed in the text -- ESP \cite[][solid]{psd13}, the $m_{\rm 200b}$ fit of T08 \cite[][dashed]{Tinker08} and the ST99 fit \cite[][short dashed]{st99} -- at three redshifts, from top to bottom $z=0.1$ (red), $z=0.3$ (blue), $z=0.7$ (black). The data points show the average mass function measured in $9$ realisations of an $N$-body simulation (see Section~\ref{sec:Nbody:sub:sims}) with the error bars representing the standard deviation of the $9$ runs. The red circles, blue triangles and black squares respectively show the measurements at $z=0.1,0.3,0.7$. \emph{(Bottom panel):} Ratios of the mass function at each redshift with the corresponding ESP mass function, with line styles and colour code as in the top panel. Dotted vertical lines in each panel show the limiting mass computed using \eqn{Moblim-Planck} for $z=0.1,0.3,0.7$ from left to right.}
\label{fig:massfunc}
\end{figure}

Figure~\ref{fig:massfunc} compares these prescriptions for the mass function. The smooth curves in the top panel show the ESP (solid), T08 (dashed) and ST99 (short dashed) mass functions for our fiducial cosmology at three redshifts -- from top to bottom, $z=0.1$ (red), $z=0.3$ (blue) and $z=0.7$ (black). For T08 we used parameter values appropriate for $m_{\rm 200b}$. The data points with error bars show the mass function measured in $N$-body simulations (described in Section~\ref{sec:Nbody:sub:sims}) performed using the same cosmological parameters. The circles, triangles and squares show measurements at redshifts $z=0.1,0.3,0.7$, respectively. For now we focus on the relative differences between the analytical mass functions, which are further highlighted in the bottom panel of the Figure which shows the ratio of the mass functions at each redshift to the corresponding ESP curve (from left to right, $z=0.7,0.3,0.1$). The horizontal dotted lines mark $10\%$ departures relative to ESP.
For illustrative purposes, the vertical dotted lines in both panels show the limiting mass from equation~\eqref{Moblim-Planck} for a Planck-like SZ survey, at the three redshifts (increasing from left to right). As we discuss below, this is not quite consistent since \eqn{Moblim-Planck} should be applied to $m_{\rm 500c}$. We show the same limits for $m_{\rm 200b}$ as well since this will be useful in building intuition regarding the results of a likelihood analysis (see Section~\ref{sec:mc:sub:fisher}). As mentioned earlier, we see that the ESP mass function agrees with T08 at the $\sim10\%$ level, except at high redshifts and masses where it substantially underpredicts the halo counts. 

One particular reason to consider the ESP framework is the natural presence of the quantities $V_\ast$ and $\gamma$. The spectral ratio $\gamma$ is related to the width of the matter power spectrum 
while $V_\ast$ is related to the typical inter-peak separation and can be thought of as a characteristic peak volume (BBKS). For power-law power spectra $P(k)\propto k^n$ with $-3<n<1$, one can prove that $\gamma$ is constant while $V_\ast\propto V$ (the exact values of the constants are not very illuminating), which means that the ESP mass fraction $f_{\rm ESP}$ for this case is explicitly universal, being a function only of the scaling variable $\nu$. For CDM-like spectra, on the other hand, $\gamma$ and $(V_\ast/V)$ both show weak but non-trivial dependencies on smoothing scale and hence mass, which means that the resulting mass function is \emph{naturally predicted} to be weakly non-universal: $f_{\rm ESP}=f_{\rm ESP}(\nu(m,z);\gamma(m),V_\ast(m))$. Although this non-universality from mass-dependence is, at first glance, quite different from the explicit \emph{redshift} dependence modelled by T08, $f_{\rm T08}=f_{\rm T08}(\nu(m,z);z)$, as seen in Figure~\ref{fig:massfunc} the ESP prediction tracks the redshift dependence of the T08 fit and $N$-body mass functions quite well. This point was first emphasized by PSD13 (see their Figure~8).

As mentioned above, the distribution $p(\beta)$ in the ESP calculation that describes the collapse barrier was chosen to simultaneously match the mass function and proto-halo overdensities of the $m_{\rm 200b}$ halos of T08. This was mainly because at the time there were no other proto-halo measurements to compare with. In principle, one should at least recalibrate $p(\beta)$ for the mass definition of interest, and possibly for additional non-universal effects. 
We will leave this for future work and, instead, throughout this paper we will use functional forms for $\der n/\der\ln m$ appropriate for $m=m_{\rm 200b}$ which are guaranteed to give a clean comparison between the ESP and T08 mass functions. The integration variable $m$ in equation~\eqref{mu-mobz} for example will then always be $m_{\rm 200b}$. To compute results appropriate for other mass definitions (in particular $m_{\rm 500c}$ which will appear later) we will explicitly model the conversion between the two mass definitions through a probability distribution, e.g. $p(\ln m_{\rm 500c}|\ln m_{\rm 200b})$, whose calibration we will discuss in detail below. Since we will treat $m_{\rm 500c}$ as an observable, this exercise will also serve as a proxy for a more realistic treatment where one would convert from, e.g., a mass function calibrated for $m_{\rm 500c}$ to a cluster observable such as $Y_{\rm SZ}$ through the distribution $p(Y_{\rm SZ}|m_{\rm 500c})$.

\section{Monte Carlo tests}   
\label{sec:montecarlo}
\noindent
In this section we assess the relative \emph{statistical} difference between the T08 and ESP mass functions, both of which as we have seen are weakly non-universal and agree with $N$-body simulations at the $\sim5$-$10\%$ level. We will do this by sampling the ESP mass function to generate a mock cluster catalog which we will analyse using the T08 mass function. Since these mock catalogs can be generated very quickly compared to a full $N$-body simulation, this comparison can be done with small statistical errors and will help us understand the role of parameter degeneracies and survey selection strategies in the presence of nonlinear systematics. This will also give us a benchmark against which to compare the results of the $N$-body analysis in Section~\ref{sec:Nbody} below. To get a rough idea of what to expect from such a comparison, we start with a Fisher analysis.

\begin{figure*}
\includegraphics[width=0.9\textwidth]{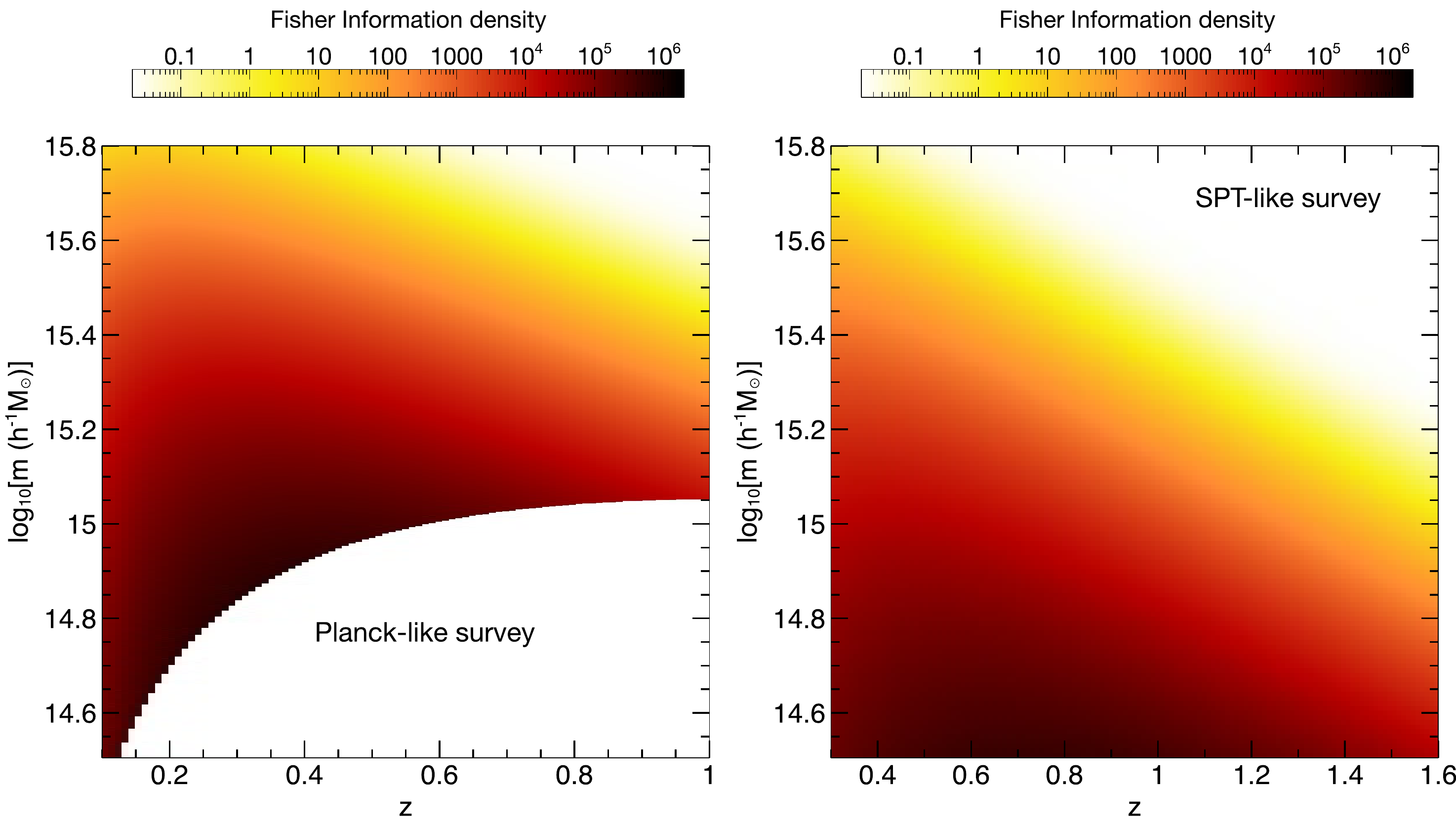}
\caption{The Fisher Information density for constraints on the single parameter $\sig_8$, defined as  $F_{ij,\sig_8\sig_8}/(\Del\log m\, \Del z)$ for bin widths $\Del\log m$ and $\Del z$ in the log-mass and redshift directions, respectively, where $F_{ij,\sig_8\sig_8}$ was defined in \eqn{fisher-clusters} and computed using the T08 $m_{\rm 200b}$ mass function for the fiducial cosmology. The \emph{left panel} shows the density for a Planck-like survey with selection threshold \eqref{Moblim-Planck} (seen as the sharp lower boundary of the coloured region), while the \emph{right panel} shows the result for an SPT-like survey with limiting mass \eqref{Moblim-SPT}. Note the different redshift ranges in the two panels. The constraints on $\sig_8$ are driven by the approximate redshift range $0.2 \lesssim z \lesssim 0.5$ for the Planck-like survey and $0.5 \lesssim z \lesssim 1.0$ for the SPT-like survey. For a fixed redshift, the constraints are driven by the smallest mass bins allowed by the selection threshold.}
\label{fig:fisherinfo}
\end{figure*}
%

\subsection{Fisher analysis}
\label{sec:mc:sub:fisher}
\noindent
The Fisher matrix is a useful tool to assess the level to which a survey can constrain a given set of parameters (see \cite{heavens09} for a review). 
Here we will use a Fisher analysis to understand which region of the $(m_{\rm ob},z)$ plane is primarily responsible for the constraints on $\sig_8$.

The Fisher matrix is defined as the expectation value (over the distribution of data) of the Hessian of the log-likelihood $\ln\Cal{L}$ with respect to parameters $\theta_a$:
\be
F_{ab} \equiv -\avg{\frac{\p^2\ln\Cal{L}}{\p\theta_a\p\theta_b}}\,.
\label{fisher-def}
\ee
For the likelihood \eqref{likelihood} appropriate for cluster cosmology, assuming that the data are drawn from a fiducial cosmology with parameter values $\theta_a^{\rm (fid)}$, the Fisher matrix reduces to \cite{hhm01}
\begin{align}
F_{ab} &= \sum_{i,j}\frac1{\mu_{ij}}\frac{\p\mu_{ij}}{\p\theta_a}\frac{\p\mu_{ij}}{\p\theta_b}
\equiv\sum_{i,j}F_{ij,ab}\,,
\label{fisher-clusters}
\end{align}
where the index $i$ runs over the bins in $m_{\rm ob}$ and $j$ over redshift bins, with all quantities being evaluated at $\theta_a = \theta_a^{\rm (fid)}$. The marginalised error on, e.g., $\sig_8$ would be the square-root of the inverse Fisher element $\sqrt{(F^{-1})_{\sig_8\sig_8}}$, while the conditional error (i.e., assuming all other parameters are held fixed) is given by $1/\sqrt{F_{\sig_8\sig_8}}$.

Considering for simplicity the case when other parameters are held fixed, it is clear from equation~\eqref{fisher-clusters} that the size of the error bar on $\sig_8$ is driven by the region in $(m_{\rm ob},z)$ where $F_{ij,\sig_8\sig_8}$ attains its maximum value. Figure~\ref{fig:fisherinfo} shows the Fisher information density (i.e., $F_{ij,\sig_8\sig_8}/(\Del\log m\, \Del z)$ for bins of width $\Del\log m$ and $\Del z$ in the log-mass and redshift directions, respectively) for our fiducial cosmology, computed using the T08 $m_{\rm 200b}$ mass function for two choices of survey selection thresholds, \eqn{Moblim-Planck} for a Planck-like survey (left panel) and \eqn{Moblim-SPT} for an SPT-like survey (right panel). Although, strictly speaking, these selection criteria apply to the $m_{\rm 500c}$ definition, the qualitative features of the mass function and hence Fisher information density should be independent of mass definition.

We see that the constraints on $\sig_8$ for a Planck-like survey are primarily driven by the redshift range $0.2 \lesssim z \lesssim 0.5$, while for an SPT-like survey the range is closer to $0.5 \lesssim z \lesssim 1.0$. Further, for any fixed redshift, the constraints are always driven by the smallest masses allowed by the selection threshold. This is sensible since the mass function is steep, so that the bins with the smallest masses always have the largest number of objects at any $z$. This feature of cluster surveys makes it especially important to accurately model mass scatter at the selection boundary.

Keeping this in mind, \fig{fig:massfunc} suggests that for a Planck-like survey the T08 mass function will systematically predict fewer objects than ESP in the relevant range of mass and redshift. Consequently, if all other parameters are held fixed, the constraint on $\sig_8$ when analysing the mock ESP catalog using the T08 mass function should be biased \emph{high} compared to the fiducial value.

\subsection{Relative bias between ESP and T08}
\label{sec:mc:sub:ESPvsTinker}
\noindent
Our basic strategy is to sample the ESP mass function using the fiducial cosmology (in particular, with the fiducial value $\sig_{\rm 8,fid}$) and generate a mock cluster catalog, adhering to a chosen selection threshold (equation~\ref{Moblim-Planck} for a Planck-like survey and equation~\ref{Moblim-SPT} for an SPT-like survey). We then analyse this catalog by computing the likelihood function \eqref{likelihood} using the T08 mass function. This will result in a posterior probability distribution $p(\sig_8)$, whose mean $\bar\sig_8$ and standard deviation $\Sigma_{\sig_8}$ can be used to quantify the level of bias between ESP and T08 by constructing the `significance' $s$ defined by
\be
s \equiv \frac{\bar\sig_8 - \sig_{\rm 8,fid}}{\Sigma_{\sig_8}}\,,
\label{significance}
\ee
which we will compute for a large number of mocks.
Ideally, the distribution of $s$ should be peaked at zero with variance close to unity (this would be exact if $s$ were Gaussian distributed with no bias). The mean or median of this distribution are then an indicator of the relative bias between the two mass functions.

Several assumptions are necessary in order to perform this comparison in practice. First, one must decide which definition of halo mass to use as the `observable' $m_{\rm ob}$. For the reasons mentioned previously, the cleanest comparison follows from using $m_{\rm 200b}$, which is what we will start with. Later, to make the analysis more realistic, we will also generate and analyse mock catalogs using $m_{\rm ob}=m_{\rm 500c}$. 
Additionally, one must choose which cosmological parameters to vary in the analysis. Ideally, this should include all parameters that are potentially degenerate with $\sig_8$. However, since the primary degeneracy of $\sig_8$ is with $\Om_{\rm m}$, we will focus on analyses where we allow only $\Om_{\rm m}$ to vary along with $\sig_8$, with several choices of priors.

\subsubsection{Results for $m_{\rm ob}\to m_{\rm 200b}$}
\label{sec:mc:sub:ESPvsTinker:subsub:m200b}
\noindent
We first discuss the case of a Planck-like survey using $m_{\rm ob}=m_{\rm 200b}$.  Specifically, we define a grid in the $(\log m_{\rm ob},z)$ plane: along the redshift direction we use central values $0.1\leq z\leq 1.0$ equally spaced with separation $\Delta z=0.05$, and along the log-mass direction we use equally spaced bins with separation $\Delta\log m=0.035$ with bin edges satisfying $\log M_{\rm ob,lim}(z) \leq \log m_{\rm ob} \leq 16$ for each $z$. The redshift range matches the one studied by the Planck Collaboration. We have checked that refining this grid and/or increasing the mass range has no effect on our results. For each bin $(m_{{\rm ob},i},z_j)$ we compute the fiducial expected number of clusters $\mu_{\rm ij}^{\rm (fid)}$ using equation~\eqref{mu-mobz} setting $\der n/\der\ln m \to \der n_{\rm ESP}/\der\ln m$ and $\avg{\ln m_{\rm ob}|\ln m}\to \ln m$ (note that the integration variable $m$ is also identified with $m_{\rm 200b}$; see the discussion at the end of Section~\ref{sec:analytical:sub:ESP})\footnote{The integral over mass in \eqn{mu-mobz} is performed over a refined mass grid, while the one over redshift is estimated using a $5$-point extended Simpson rule.}. We assume a $10\%$ Lognormal error in mass estimation, setting $\sig_{\ln m_{\rm ob}} \to 0.5\ln(1.1/0.9) \simeq 0.1$. 

For the limiting mass $M_{\rm ob,lim}(z)$ we use equation~\eqref{Moblim-Planck}. This is not quite consistent, since that relation is appropriate for the $m_{\rm 500c}$ definition rather than $m_{\rm 200b}$. 
However, since $m_{\rm 500c} < m_{\rm 200b}$ for any object (see, e.g., Appendix~\ref{app:masscal}), this is a simple way of mimicking the effect of an increased number of clusters due to longer integration time as is expected for near-future analyses of the complete Planck data set. We discuss this further below. We find that the total fiducial expected number of clusters using the ESP mass function is $\sim1150$ in this case.

\begin{figure}
\includegraphics[width=0.45\textwidth]{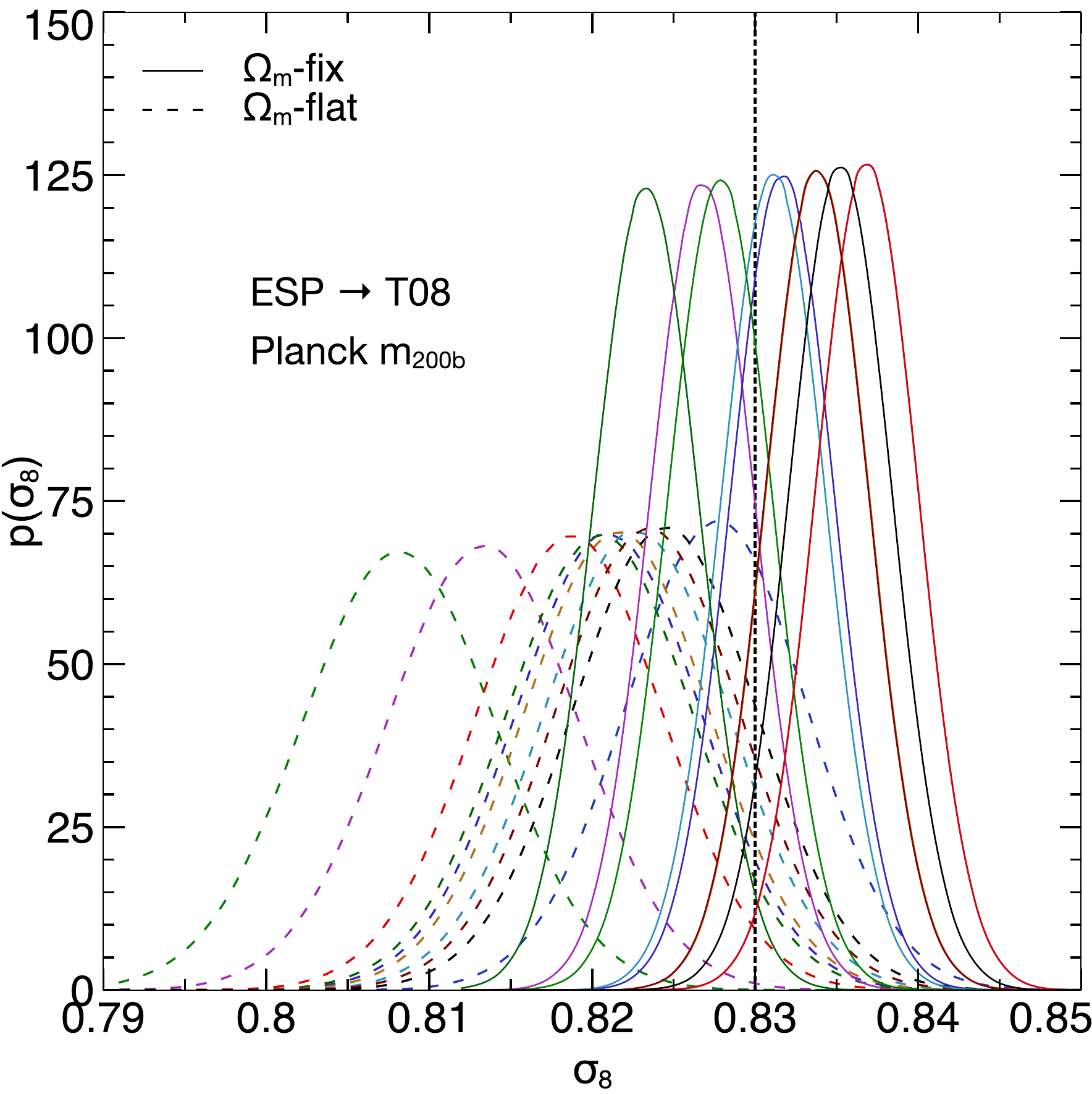}
\caption{Randomly chosen posterior distributions $p(\sig_8)$ from the likelihood analysis of mock catalogs generated using the ESP mass function with the fiducial cosmology (vertical dotted line shows the input value of $\sig_8$) and analysed using the T08 mass function for the $m_{\rm 200b}$ mass definition, when $\Om_{\rm m}$ is fixed at its true value (solid curves) and marginalised over a broad, flat prior (dashed curves). Each curve is approximately symmetric around its peak value, a feature that is shared by all the posterior distributions we analyse (not shown).}
\label{fig:psig8}
\end{figure}

\begin{figure*}
\includegraphics[width=0.9\textwidth]{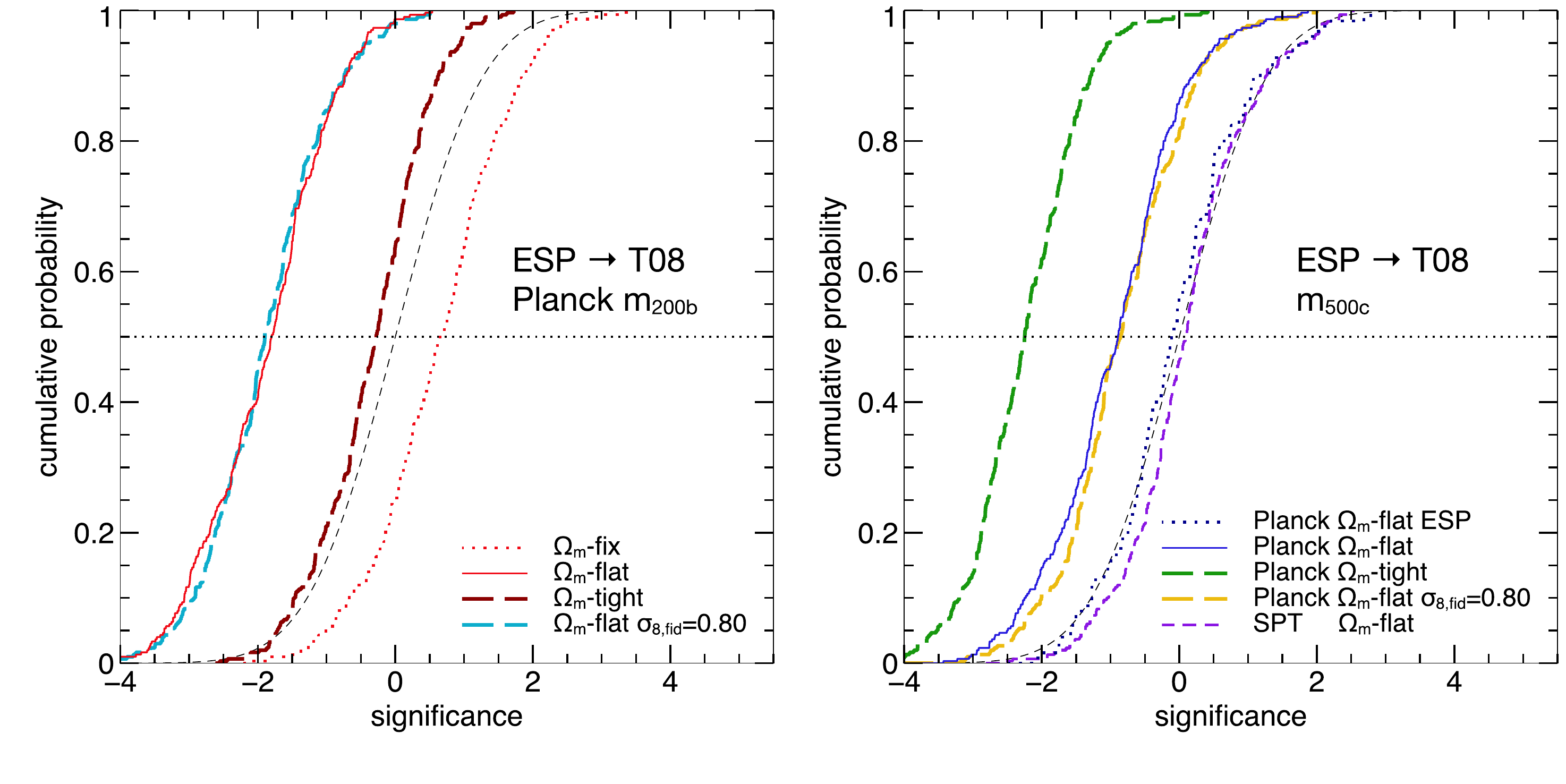}
\caption{Cumulative distributions of the significance $s$ (equation~\ref{significance}) of the bias in $\sig_8$ for the likelihood analysis of $N$ mock catalogs generated using the ESP mass function with the fiducial cosmology and analysed using the T08 mass function for various choices of survey selection threshold, mass definition and prior on $\Om_{\rm m}$. 
Unless specified, we set $N=300$.
The thin dashed black line in each panel shows the Gaussian distribution for comparison. \emph{(Left panel):} Mass definition $m_{\rm 200b}$, Planck-like selection threshold \eqref{Moblim-Planck}. The bias on $\sig_8$ moves to more negative values as the prior on $\Om_{\rm m}$ is relaxed from a fixed value (dotted red) to a tight $2\%$ Gaussian (thick long-dashed brown) to a broad flat distribution (solid red). The result is independent of the fiducial value of $\sig_8$, as shown by the distribution for $\sig_{\rm 8,fid}=0.80$ (thick long-dashed blue). \emph{(Right panel):} Mass definition $m_{\rm 500c}$, with catalogs generated and analysed using the mass calibration scheme $C_1$ of Appendix~\ref{app:masscal}. This time, the bias for a Planck-like survey when using a tight prior on $\Om_{\rm m}$ (thick long-dashed green) is more negative than for a flat prior (solid blue). As before, the result is independent of the fiducial value of $\sig_8$, as shown by the distribution for $\sig_{\rm 8,fid}=0.80$ (thick long-dashed yellow). The bias for an SPT-like survey (selection threshold~\ref{Moblim-SPT}) with a flat $\Om_{\rm m}$ prior (short-dashed purple) is considerably less significant than the corresponding bias for a Planck-like survey.
As a sanity check, the likelihood analysis using the ESP mass function for the Planck-like survey leads to unbiased results (dotted blue, $N=125$).}
\label{fig:cumulative}
\end{figure*}

Having computed $\mu_{ij}^{\rm (fid)}$, we generate a mock catalog by drawing a Poisson random number $N_{ij}$ with mean $\mu_{ij}^{\rm (fid)}$ for each bin $(m_{{\rm ob},i},z_j)$. 
The next step is to compute the likelihood \eqref{likelihood} for this data set $\{N_{ij}\}$ for arbitrary values of the cosmological parameters, with $\mu_{ij}$ computed exactly as above, except that we use the T08 mass function $\der n/\der\ln m \to \der n_{\rm T08}/\der\ln m$. To begin with, we do this allowing only $\sig_8$ to vary, keeping $\Om_{\rm m}$ and all other parameters fixed at their fiducial values. Assuming a broad, flat prior on $\sig_8$, the likelihood for each mock data set $\{N_{ij}\}$ leads to a normalised posterior distribution for $\sig_8$: $p(\sig_8|\{N_{ij}\}) = \Cal{L}(\{N_{ij}\}|\sig_8)/\int\der\sig_8\,\Cal{L}(\{N_{ij}\}|\sig_8)$. We repeat this analysis $N$ times; unless specified, we use $N=300$ hereafter. The solid curves in Figure~\ref{fig:psig8} show $p(\sig_8)$ for $10$ randomly chosen mocks. The prior on $\sig_8$ is always chosen to be broad enough to comfortably envelope the likelihood.
The result for the cumulative distribution of $s$ (equation~\ref{significance}) is shown as the dotted red curve in the left panel of Figure~\ref{fig:cumulative}. (For comparison, the thin dashed black curve shows the cumulative Gaussian distribution.) As anticipated in Section \ref{sec:mc:sub:fisher}, the T08 mass function predicts a value of $\sig_8$ that is biased high compared to the fiducial; the median significance can be read off the Figure and the mean significance is $\avg{s} = +0.62\pm0.06$. 

Next, we allow the value of $\Om_{\rm m}$ to vary within a flat prior $0.1 \leq \Om_{\rm m} \leq 0.6$ simultaneously with $\sig_8$, keeping all other parameters fixed. We have checked that increasing the range of the prior does not affect our results. In this case the posterior distribution of $\sig_8$ for a given mock data set is computed by marginalising the likelihood over $\Om_{\rm m}$:
\be
p(\sig_8|\{N_{ij}\}) = \frac{\int\der\Om_{\rm m}\,p(\Om_{\rm m})\,\Cal{L}(\{N_{ij}\}|\sig_8,\Om_{\rm m})}{\int\der\sig_8\int\der\Om_{\rm m}\,p(\Om_{\rm m})\,\Cal{L}(\{N_{ij}\}|\sig_8,\Om_{\rm m})}\,.
\label{psig8}
\ee
We perform the necessary integrals on a grid in $\Om_{\rm m}$ and $\sig_8$. The dashed curves in Figure~\ref{fig:psig8} show the posterior $p(\sig_8)$ for $10$ of the $300$ mocks, and the overall cumulative distribution of $s$ is shown as the solid red curve in the left panel of Figure~\ref{fig:cumulative}. We see a dramatic difference in the latter as compared to the case when $\Om_{\rm m}$ was held fixed; the T08 mass function now predicts a value of $\sig_8$ that is biased significantly \emph{low} compared to the fiducial, with a mean significance $\avg{s} = -1.89\pm0.05$. This behaviour is due to the strong degeneracy between $\sig_8$ and $\Om_{\rm m}$ coupled with the systematic differences between the two mass functions\footnote{Similar effects can be seen in much simpler cases as well; e.g., consider fitting the systematically biased model $y=mx+c+0.05x^2$ to data drawn from the true model $y_{\rm true}=x$ with errors $\sig_y=0.01$. If the fit is performed fixing $m=1$, the best fit value of $c$ would be negative. However, if $m$ and $c$ both vary, positive best-fit values of $c$ are easily possible, especially if the range of $x$ is restricted to values far from $x=0$.}.

As an intermediate example to the two extreme $\Om_{\rm m}$ priors discussed above, we consider the case of a tight Gaussian prior with a width of $2\%$ of the fiducial value. The cumulative distribution of $s$ in this case is shown as the thick long-dashed brown curve in the left panel of Figure~\ref{fig:cumulative}, which lies between the fixed and flat prior cases with a mean significance $\avg{s}=-0.35\pm0.05$. Finally, as a check that our results are independent of the fiducial values of the parameters, we repeat the analysis using the flat prior on $\Om_{\rm m}$ for a fiducial cosmology identical to the previous, except that we use $\sig_{8{,\rm fid}}=0.80$. The resulting distribution of $s$ is shown as the thick long-dashed cyan curve in the left panel of Figure~\ref{fig:cumulative}; this is very close to the corresponding curve for $\sig_{8{,\rm fid}}=0.83$ and has a mean value $\avg{s}=-1.88\pm0.05$.

\subsubsection{Results for $m_{\rm ob}\to m_{\rm 500c}$}
\label{sec:mc:sub:ESPvsTinker:subsub:m500c}
\noindent
To make the analysis more realistic, we now consider the case where we ``observe'' the mass $m_{\rm 500c}$ for the clusters, while the mass functions still predict counts for $m_{\rm 200b}$. We can model this situation using equation~\eqref{mu-mobz} by retaining the identification $m=m_{\rm 200b}$ for the integration variable as before, but using $m_{\rm ob}=m_{\rm 500c}$. This means we must model the distribution $p(\ln m_{\rm 500c}|\ln m_{\rm 200b})$, which we discuss next. Notice, however, that this means we are modeling a stochastic proxy $m_{\rm 500c}$ in place of the  `true' mass $m_{\rm 200b}$, and that this proxy is strongly biased since we always have $m_{\rm 500c}<m_{\rm 200b}$ (e.g., Figure~\ref{fig:masscalC1}). This exercise is therefore quite close in spirit to more realistic analyses involving scaling relations of biased mass proxies.

\begin{figure*}
\includegraphics[width=0.9\textwidth]{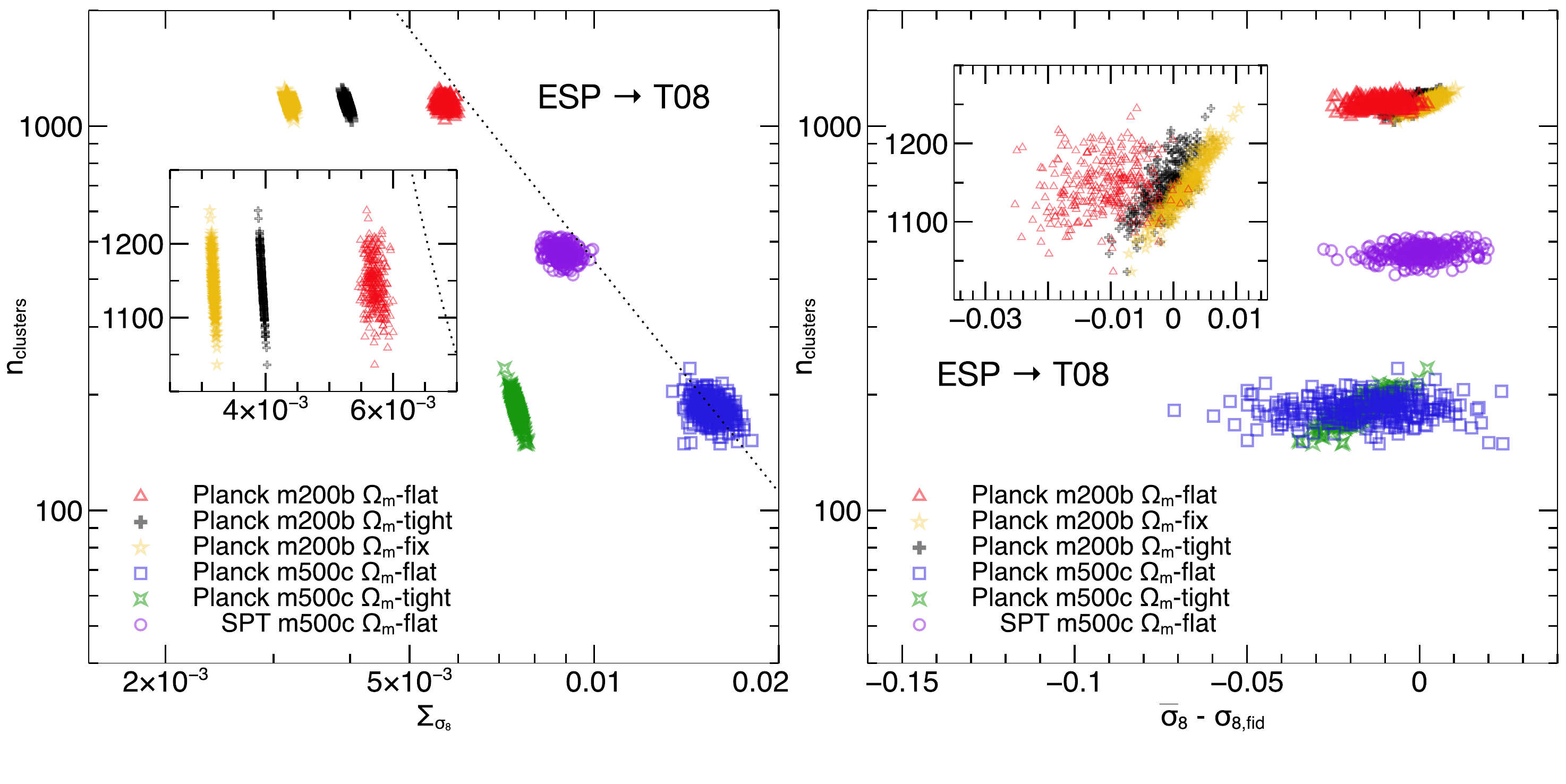}
\caption{Joint distributions of the number of mock clusters $n_{\rm clusters}$ with the standard deviation $\Sigma_{\sig_8}$ of the posterior $p(\sig_8)$ \emph{(left panel)} and with the absolute bias $\bar\sig_8-\sig_{\rm 8,fid}$ \emph{(right panel)} as measured in the Monte Carlo mock catalogs discussed in Figure~\ref{fig:cumulative}. Results are displayed for different survey selection thresholds (Planck/SPT), mass definitions ($m_{\rm 200b}$/$m_{\rm 500c}$) and $\Om_{\rm m}$ priors (flat/tight/fixed) as marked. The insets zoom in on the $m_{\rm 200b}$ results using a linear scale on both axes. The dotted curve in the left panel shows the behaviour $\avg{n_{\rm clusters}}^{-1/2}$ normalised to the Planck $m_{\rm 500c}$ flat-$\Om_{\rm m}$ cloud. The trends in these panels help understand the behaviour of the cumulative distributions of Figure~\ref{fig:cumulative} (see text for a discussion).}
\label{fig:ncdist}
\end{figure*}

For any given dark matter halo, the relation between different SO mass definitions can be derived if we know the density profile of dark matter in the halo. While this usually requires numerical inversions of integrals over the profile, in the case of the \citet*[][NFW]{nfw96} profile, 
\citet{hk03} have provided an accurate analytical prescription which we adopt here. The conversion requires knowledge of the `concentration' parameter which governs the shape of the NFW profile; this is a stochastic quantity whose distribution must be calibrated from simulations. The stochasticity in the concentration leads to a scatter in the $m_{\rm 500c}$-$m_{\rm 200b}$ relation, and hence determines the distribution $p(\ln m_{\rm 500c}|\ln m_{\rm 200b})$ which we approximate as being Gaussian in $\ln(m_{\rm 500c})$. We describe our procedure for calibrating this conditional distribution in detail in Appendix~\ref{app:masscal}. 

This conversion necessarily depends on the value of $\Om_{\rm m}$, and modeling this dependence accurately is therefore essential in obtaining unbiased cosmological constraints. In this Section we use the calibration scheme $C_1$ described in Appendix~\ref{app:masscal}. Briefly, this fixes a cosmology-independent scatter (see below) and a cosmology-dependent mean for the distribution $p(\ln m_{\rm 500c}|\ln m_{\rm 200b})$. The mean value $\avg{\ln m_{\rm 500c}|\ln m_{\rm 200b}}$ follows from the analytical calculation mentioned above and assumes a specific form for the mean of the stochastic  concentration-mass-redshift relation (equation~\ref{app-meanc200b} with the $C_1$ parameters from equation~\ref{app-calschemes}).
This is sufficient for measuring the relative bias between ESP and T08 since we use the same scheme to generate and analyse the clusters. Later, when we consider cluster catalogs built from $N$-body simulations, the absolute calibration of this relation will become important, and we will explore the consequences of changing the mass calibration at the $\sim5\%$ level.

Knowing $p(\ln m_{\rm 500c}|\ln m_{\rm 200b})$ for any cosmology, we proceed in a manner similar to that used in Section~\ref{sec:mc:sub:ESPvsTinker:subsub:m200b}, with the following changes. For bins in redshift and $\log m_{\rm ob} = \log m_{\rm 500c}$ we compute the expected number of clusters in the fiducial cosmology $\mu_{ij}^{\rm (fid)}$ using equation~\eqref{mu-mobz}, with $p(\ln m_{\rm 500c}|\ln m_{\rm 200b})$ evaluated for the fiducial cosmology. For the Planck-like survey we also introduce an additional $10\%$ scatter in quadrature, which is an easy way of accounting for realistic scatter in the mass-observable relation used for actual data but assumes that there is no further mass bias\footnote{The choice of $10\%$ is motivated by the typical scatter of mass-observable scaling relations (see, e.g., Appendix A.3.3 of \cite{planck13-XX-SZcosmo}).}.
Namely, we set the width $\sig_{\ln m_{\rm ob}}\to\sqrt{\sig_{\ln m_{\rm 500c}}^2 + \sig_{\ln m,{\rm error}}^2}$, where $\sig_{\ln m_{\rm 500c}}=0.18$ is the width of the $\ln m_{\rm 500c}$-$\ln m_{\rm 200b}$ relation calibrated in Appendix~\ref{app:masscal} and, as before, $\sig_{\ln m,{\rm error}} = 0.5\ln(1.1/0.9)\simeq0.1$, so that $\sig_{\ln m_{\rm ob}}=0.21$.
The catalog of number counts $\{N_{ij}\}$ is generated as before by Poisson-sampling $\mu_{ij}^{\rm (fid)}$; the total expected number of clusters in this case is $\sim184$, close to that in the current release of Planck SZ clusters. The likelihood analysis uses equation~\eqref{mu-mobz} again, with the mass function as well as $p(\ln m_{\rm 500c}|\ln m_{\rm 200b})$ evaluated on a grid of $\Om_{\rm m}$ and $\sig_8$ values. 

The analysis for the cumulative distribution of the significance $s$ (equation~\ref{significance}) proceeds exactly as before, and the results are shown in the right panel of Figure~\ref{fig:cumulative}. Due to the added complexity of calibrating the $m_{\rm 500c}$-$m_{\rm 200b}$ relation, as a sanity check we first perform the likelihood analysis using the ESP mass function itself, which should lead to an unbiased result. As shown by the dotted blue curve, this is indeed the case; the distribution of $s$ (for $N=125$ in this case) and a flat prior on $\Om_{\rm m}$ is close to Gaussian (the latter is shown again as the thin dashed black curve) with a mean value $\avg{s} = -0.08\pm0.09$. (We used fewer mocks since the evaluation of the ESP mass function at present is quite time-consuming.)

We next analyse the mock ESP data with the T08 mass function. This time, we see a qualitative difference in the relative trends of the distribution of $s$ as the prior on $\Om_{\rm m}$ is changed. The distribution for a flat prior is shown by the blue solid line and has $\avg{s}=-0.96\pm0.06$. The distribution for a tight $2\%$ Gaussian prior on the other hand is shown by the thick long-dashed green curve and has a \emph{more negative} bias with $\avg{s}=-2.25\pm0.04$. (We have also checked that using a $10\%$ Gaussian prior gives results that are between the $2\%$ and flat cases.) We discuss this further below.

As before, we have checked that the results are independent of the fiducial value of $\sig_8$; the thick long-dashed yellow curve shows the distribution of $s$ for a flat prior on $\Om_{\rm m}$ when using $\sig_{8,{\rm fid}}=0.80$. This is close to the corresponding curve for $\sig_{8,{\rm fid}}=0.83$ and has $\avg{s}=-0.85\pm0.05$.

\begin{table*}
\centering
\begin{tabular}{|c|ccc|c|c|}
\hline
 & \multicolumn{4}{c|}{Planck} & SPT \\
\hline
 & \multicolumn{3}{c|}{$\sig_{8,{\rm fid}}=0.83$} & $\sig_{8,{\rm fid}}=0.80$ & $\sig_{8,{\rm fid}}=0.83$\\
\hline
mass def. & flat-$\Om_{\rm m}$ & tight-$\Om_{\rm m}$ & fixed-$\Om_{\rm m}$ & flat-$\Om_{\rm m}$ & flat-$\Om_{\rm m}$ \\
\hline\hline
$m_{\rm 200b}$  & $-1.89\pm0.05$ & $-0.35\pm0.05$ & $+0.62\pm0.06$ & $-1.88\pm0.05$ & -- \\[1pt]
$m_{\rm 500c}$  & $-0.96\pm0.06$ & $-2.25\pm0.04$ & -- & $-0.85\pm0.05$ & $+0.09\pm0.05$ \\[1pt]
\hline
\end{tabular}
\caption{Summary of mean values of the significance $\avg{s}$ for the Monte Carlo mock catalogs generated using the ESP mass function and analysed using the T08 mass function, whose full distributions were displayed in Figures~\ref{fig:cumulative} and~\ref{fig:ncdist}. The rows correspond to two different mass definitions while the columns indicate different priors on $\Om_{\rm m}$, values of $\sig_{\rm 8,fid}$ and survey selection threshold. The error bars are estimates of the standard error on the mean from $N$ mock catalogs ($N=300$ for all cases), and are numerically always close to $1/\sqrt{N}$. See text for a discussion of the trends.}
\label{tab:mocks}
\end{table*}

Finally, we also repeated this analysis for an SPT-like survey, using the selection threshold \eqref{Moblim-SPT} and $f_{\rm sky}=0.06$. In this case we add a $20\%$ mass error in quadrature to the scatter of the $m_{\rm 500c}$-$m_{\rm 200b}$ relation (which is otherwise treated identically to the Planck case), setting $\sig_{\ln m_{\rm ob}}=0.27$ in total. The total expected number of clusters in the fiducial cosmology is $\sim470$. The resulting distribution of $s$ for a flat prior on $\Om_{\rm m}$ is shown by the short-dashed purple line in the right panel of Figure~\ref{fig:cumulative} and has $\avg{s}=+0.09\pm0.05$, considerably smaller in magnitude than the corresponding value for the Planck-like survey.

We explore the behaviour of the relative bias further in Figure~\ref{fig:ncdist}, which shows the joint distribution of the total number of clusters $n_{\rm clusters}$ drawn in our mocks with the r.m.s. $\Sigma_{\sig_8}$ of the posterior (left panel) and with the absolute bias $\bar\sig_8-\sig_{8,{\rm fid}}$ (right panel). The different clouds of points show the results for the Planck-$m_{\rm 200b}$ analysis with fixed, tight and flat $\Om_{\rm m}$ priors (these have $\sim1150$ clusters on average), the Planck-$m_{\rm 500c}$ analysis with tight and flat $\Om_{\rm m}$ priors ($\sim184$ clusters on average), and the SPT-$m_{\rm 500c}$ analysis for a flat $\Om_{\rm m}$ prior  ($\sim470$ clusters on average). The insets in each panel zoom in on the $m_{\rm 200b}$ distributions, with a linear scale on each axis.

In the left panel, we see that the typical standard deviation $\Sigma_{\sig_8}$ behaves as expected. For the same prior on $\Om_{\rm m}$ it decreases approximately like $\avg{n_{\rm clusters}}^{-1/2}$ (the dotted curve shows this relation normalised to the Planck-$m_{\rm 500c}$ case with a flat $\Om_{\rm m}$ prior), with only small systematic effects due to the choice of mass definition and survey selection as can be seen from the small shifts in the SPT-$m_{\rm 500c}$ and Planck-$m_{\rm 200b}$ clouds relative to the dotted line.
For a fixed survey and mass definition, on the other hand, the typical $\Sigma_{\sig_8}$ decreases as the prior on $\Om_{\rm m}$ is tightened. 

The right panel of Figure~\ref{fig:ncdist} shows more interesting behaviour. We see that the Planck-$m_{\rm 200b}$ clouds behave as expected from the $\Om_{\rm m}$-$\sig_8$ degeneracy (see the discussion in Section~\ref{sec:mc:sub:ESPvsTinker:subsub:m200b}), with a typical absolute bias $\bar\sig_8-\sig_{8,{\rm fid}}$ that increases from negative to positive values as the prior on $\Om_{\rm m}$ is tightened. The Planck-$m_{\rm 500c}$ clouds, on the other hand, have approximately the same typical values of $\bar\sig_8-\sig_{8,{\rm fid}}$, although with very different scatters. This change in behaviour is not surprising given the $\Om_{\rm m}$ dependence of the mass conversion, which alters the $\Om_{\rm m}$-$\sig_8$ degeneracy. Due to the numerical complexity of the problem, it is difficult to make a more precise statement. Combined with the behaviour of $\Sigma_{\sig_8}$ in the left panel, however, this explains the reversal of trend of the distribution of $s$ in the $m_{\rm 500c}$ case as compared to $m_{\rm 200b}$ that was seen in Section~\ref{sec:mc:sub:ESPvsTinker:subsub:m500c}.

Finally, the SPT-$m_{\rm 500c}$ cloud has a substantially different typical value of $\bar\sig_8-\sig_{8,{\rm fid}}$ as compared to the Planck-$m_{\rm 500c}$ clouds, showing that the survey selection threshold plays an important role in determining the overall effect of nonlinear systematics. This is sensible since different selection thresholds explore different regimes of the mass function in the $m_{\rm ob}$-$z$ plane.

To summarize this Section, we have explored the sensitivity of the relative bias between the ESP and T08 mass functions when constraining $\sig_8$ to the number of clusters observed by the survey, the nature of the prior on parameters degenerate with $\sig_8$ (we focused on $\Om_{\rm m}$), and the form of the survey selection threshold. Table~\ref{tab:mocks} summarizes these results. 

\section{$N$-body tests}   
\label{sec:Nbody}
\noindent
So far we have only studied the relative bias between two analytical mass functions. To properly assess the level of absolute bias inherent in either of them, we repeat the analysis of the previous section replacing the mock catalogs with halos identified in $N$-body simulations. 

One might argue that the T08 mass function is a fit to simulations in the first place, so it must be unbiased by construction. One must keep in mind, however, that the best-fit parameters reported by T08 are generally used \emph{without} accounting for their error covariance matrix \cite{rozo+10,vanderlinde+10,benson+13,hasselfield+13,reichardt+13,planck13-XX-SZcosmo}. 
Further, observables such as $Y_{\rm 500}$ for SZ surveys \cite{planck13-XX-SZcosmo} and $Y_X$ for X-ray surveys \cite{kvn06} have been shown to correlate well with $m_{\rm 500c}$, whereas the T08 fits are for $m_{\Delta{\rm b}}$. 
The standard practice has therefore been to interpolate between the T08 fits setting $\Delta = 500/\Om_{\rm m}(z)$.
Ignoring parameter errors is similar to ignoring the scatter in the $m_{\rm 500c}$-$m_{\rm 200b}$ relation discussed earlier, which can have substantial effects near the threshold mass of the survey. Since this is the region which drives the parameter constraints (c.f. Section~\ref{sec:mc:sub:fisher}), we believe it is worth performing a detailed comparison between the T08 fitting function and mock cluster catalogs built upon $N$-body simulations\footnote{We note that \citet{rozo+07} have demonstrated that using the \citet{jenkins+01} mass function fit for optically selected clusters in the Sloan Digital Sky Survey without marginalising over the fit parameters should lead to unbiased constraints on $\sig_8$. The observable in this case is the cluster richness which is calibrated directly against $m_{\rm 200b}$.}. A similar analysis is also useful for the ESP mass function which has far less input from simulations than T08 as discussed previously, and which has not been subjected to statistical tests of this nature to date.

We describe our simulations and the resulting catalogs below and then repeat the likelihood analysis of the previous Section, this time analysing the $N$-body based catalogs using both the T08 and ESP mass functions.

\subsection{Simulations and mock catalogs}   
\label{sec:Nbody:sub:sims}
\noindent
We have run $N$-body simulations of cold dark matter in a periodic cubic box of comoving size $L_{\rm box}=2h^{-1}$Gpc using the tree-PM code\footnote{http://www.mpa-garching.mpg.de/gadget/} \textsc{Gadget-2} \cite{springel:2005}. The cosmology was the same as the fiducial one used above, with a transfer function computed using the \citet{eh98} prescription (see Section~\ref{sec:intro} for the parameter values). We used $N_{\rm part}=1024^3$ particles, giving a particle mass of $m_{\rm part}=6.5\times10^{11}\Ms$, and the force resolution was set to $\epsilon = 65h^{-1}$kpc comoving ($1/30$ of the mean particle separation), with a $2048^3$ PM grid. Initial conditions were generated at $z=99$ employing $2^{\rm nd}$-order Lagrangian Perturbation Theory \cite{scoccimarro98}, using the code\footnote{http://www.phys.ethz.ch/$\sim$hahn/MUSIC/} \textsc{Music} \cite{hahn11-music} in single-resolution mode where it generates a realisation of Gaussian density fluctuations in Fourier space with the chosen matter power spectrum.
We ran $9$ realisations of this simulation by changing the random number seed used for generating the initial conditions. The simulations were run on the Brutus cluster\footnote{http://www.cluster.ethz.ch/index\_EN} at ETH Z\"urich.

The above settings imply that the smallest halo masses needed to generate a Planck-like catalog are resolved with $\sim400$ particles\footnote{This is true for each mass definition we consider. That is, our smallest $m_{\rm \Delta}$ will be resolved with $\sim400$ particles inside $R_\Delta$ for both $\Delta=200$b as well as $\Delta=500$c. Of course, these cases would correspond to very different halos.}. We discuss mass resolution effects in Appendix~\ref{app:restest}. To identify halos, we used the code\footnote{http://code.google.com/p/rockstar/} \textsc{Rockstar} \cite{behroozi13-rockstar}, which assigns particles to halos based on an adaptive hierarchical FoF algorithm in $6$-dimensional phase space. 
\textsc{Rockstar} has been shown to be robust for a variety of diagnostics such as density profiles, velocity dispersions and merger histories, and the resulting mass function agrees with T08 at the few per cent level at $z=0$ \cite{behroozi13-rockstar} (see also below). 
A convenient aspect of \textsc{Rockstar} is that by default it outputs a number of values of mass for a single object, including values of $m_{\rm 200b}$ and $m_{\rm 500c}$ which we are interested in here. We only consider parent halos in this work and ignore subhalos as independent objects. The masses reported by \textsc{Rockstar}, however, account for the total mass of each object which includes the mass contained in any subhalos, and are therefore appropriate for our purpose.

\begin{table*}
\centering
\begin{tabular}{|c|c|c|cc|cc|c|cc|}
\hline
& & \multicolumn{5}{c|}{T08} & \multicolumn{3}{c|}{ESP} \\
\hline
& & $m_{\rm 200b}$ & \multicolumn{2}{c|}{$m_{\rm 500c}$} & \multicolumn{2}{c|}{$m_{\rm 500c}$ {\scriptsize (no scatter)}} & $m_{\rm 200b}$ & \multicolumn{2}{c|}{$m_{\rm 500c}$}\\
\hline
selection & $\Om_{\rm m}$ prior &  & $C_1$ & $C_2$ & $C_1$ & $C_2$ & & $C_1$ & $C_2$ \\
\hline\hline
Eqn.~\eqref{Moblim-Planck} & flat   & $+1.6\pm0.5$ & $+0.2\pm0.4$ & $+0.5\pm0.4$ & $+1.7\pm0.4$ & $+2.0\pm0.4$ & $+3.2\pm0.5$ & $+0.6\pm0.4$ & $+1.0\pm0.4$ \\[1pt]
& fixed  & $+0.5\pm0.3$ & $-1.0\pm0.3$ & $+0.3\pm0.3$ & -- & -- & $-0.2\pm0.3$ & $-1.9\pm0.3$ & $-0.5\pm0.3$ \\[1pt]
\hline
Eqn.~\eqref{Moblim-PlanckII} & flat   & $+1.9\pm0.5$ & $+0.4\pm0.3$ & $+0.8\pm0.3$ & $+2.1\pm0.3$ & $+2.5\pm0.3$ & -- & -- & -- \\[1pt]
\hline
\end{tabular}
\caption{Summary of mean values of the significance $\avg{s}$ for the $N$-body based mock catalogs discussed in Section~\ref{sec:Nbody}, with errors given by the standard error over $9$ independent realisations. The mocks were analysed using the T08 \emph{(left block)} and ESP \emph{(right block)} mass functions, for two choices of prior on $\Om_{\rm m}$ (flat and fixed) and for two choices of mass definition ($m_{\rm 200b}$ and $m_{\rm 500c}$). Further, for $m_{\rm 500c}$, results are shown for two choices of mass calibration scheme $C_1$ and $C_2$ as described in Appendix~\ref{app:masscal}. Results are also shown for the $m_{\rm 500c}$ analysis with T08 for both schemes when the intrinsic scatter of the distribution $p(\ln m_{\rm 500c}|\ln m_{\rm 200b})$ is artificially set to zero (columns labelled ``no scatter'') which is similar to ignoring the errors on the T08 parameter values as is routinely done in cluster likelihood analyses. The last row gives the results for the T08 analysis with a flat $\Om_{\rm m}$ prior when using \eqn{Moblim-PlanckII} to define the selection threshold instead of \eqn{Moblim-Planck}. See text for further discussion.}
\label{tab:Nbody}
\end{table*}

For each realisation, we stored density snapshots at $10$ equally spaced redshifts in the range $0.1\leq z\leq 1.0$ appropriate for a Planck-like survey. These snapshots were used to construct $9$ lightcones, each spanning $f_{\rm sky}=0.235$, as described in Appendix~\ref{app:lightcones}. We use the $m_{\rm 200b}$ and $m_{\rm 500c}$ masses for the halos recorded by \textsc{Rockstar} as our mass proxies and add a $10\%$ Lognormal scatter in each case to model mass uncertainties as we did for the Monte Carlo mocks. Upon using \eqn{Moblim-Planck} for the survey selection threshold, this gives us $847$ clusters on average for the $m_{\rm 200b}$ case and $137$ clusters on average for $m_{\rm 500c}$.
Since the redshift bins are quite thick ($\Del z=0.1$) and we only use a single snapshot per bin to obtain halos, we are effectively approximating the mass function as being piece-wise constant across bins. We therefore alter our likelihood analysis as described below in order to be consistent with this approximation.

The points in Figure~\ref{fig:massfunc} show the $m_{\rm 200b}$ mass function of the \textsc{Rockstar} halos, averaged over the nine realisations, at three redshifts $z=0.1$ (red circles), $z=0.3$ (blue triangles) and $z=0.7$ (black squares). The error bars show the standard deviation in each bin over the $9$ realisations. We see from the bottom panel that the measured mass function agrees with the T08 analytic form to within a few per cent at all but the highest redshifts where the agreement is at the $\sim10\%$ level. The low redshift results are consistent with those reported by \citet{behroozi13-rockstar} at $z=0$. The ESP mass function also provides a good description of the $N$-body measurements, agreeing at $\lesssim10\%$ for low redshifts and at about $20\%$ at higher redshifts.

\subsection{Likelihood Analysis}
\label{sec:Nbody:sub:analysis}
\noindent
In order to compare apples with apples, we have modified the likelihood analysis of the previous Section as follows. Since our $N$-body ``lightcones'' were actually constructed by approximating the numerical mass function as being piece-wise constant over redshift bins of width $\Delta z=0.1$ (see Appendix~\ref{app:lightcones} for details), we \emph{analyse} the resulting cluster catalogs in exactly the same way. Namely, we replace the expected cluster count $\mu_{ij}$ in equation~\eqref{mu-mobz} by
\begin{align}
\mu_{ij}^{\rm (LC)} &= f_{\rm sky}\Delta V_j\int\der\ln m \frac{\der n}{\der\ln m}(m,z_j)\notag\\
&\ph{\int}\times 
\frac12\bigg[\erf{\frac{\avg{\ln m_{\rm ob}|\ln m} - \ln m_{{\rm ob},i}^{-}}{\sqrt{2}\sig_{\ln m_{\rm ob}}}}\notag\\
&\ph{f\int\frac12} 
 - \erf{\frac{\avg{\ln m_{\rm ob}|\ln m} - \ln m_{{\rm ob},i}^{+}}{\sqrt{2}\sig_{\ln m_{\rm ob}}}}\bigg]\,,
\label{mu-mobz-LC}
\end{align}
where 
\be
\Delta V_j = \frac{4\pi}{3}\left(r_{\rm com}(z_j^{+})^3 - r_{\rm com}(z_j^{-})^3\right)\,,
\label{DeltaV}
\ee
with $r_{\rm com}(z) = \int_0^z\der z'\,H(z')^{-1}$ the cosmology-dependent comoving distance to redshift $z$, and where $z_j^{\pm} = z_j\pm\Delta z/2$ are the bin edges, with $0.1\leq z_j \leq 1.0$ and $\Delta z=0.1$, and we set $f_{\rm sky}=0.235$ to be consistent with the simulation.

The rest of the likelihood analysis proceeds as before. For both T08 and ESP, we analyse the likelihood using either a fixed value or flat prior for $\Om_{\rm m}$ and using either $m_{\rm 200b}$ or $m_{\rm 500c}$ as the mass proxy (in each case accounting for the $10\%$ mass error introduced when constructing the catalog.) 
In the present case, we also explore an additional effect, which is the systematic error in the $m_{\rm 500c}$-$m_{\rm 200b}$ calibration of Appendix~\ref{app:masscal}. For the Monte Carlo analysis of the previous Section, we had only used the calibration scheme $C_1$ of Appendix~\ref{app:masscal}. This was sufficient since we were then only interested in the relative difference between ESP and T08. With the $N$-body based catalogs, however, the absolute calibration becomes important. We therefore also analyse the catalogs assuming the calibration scheme $C_2$ in Appendix~\ref{app:masscal}. Note that these two mass calibrations differ only at the $\sim5\%$ level.

To mimic the effect of ignoring parameter errors when using the T08 fits, as is routinely done, we analyse the $m_{\rm 500c}$ case for the flat $\Om_{\rm m}$ prior after artificially setting the intrinsic scatter in $p(\ln m_{\rm 500c}|\ln m_{\rm 200b})$ to zero and only keeping the $10\%$ mass error mentioned earlier. We expect that this treatment is more extreme than the interpolation of the T08 fits that is normally used, since the interpolation likely gives a better handle on the mean mass calibration than our single jump from $m_{\rm 200b}$ to $m_{\rm 500c}$, although this issue deserves a more thorough investigation. In any case, the results of this `no scatter' analysis should at least serve as useful upper bounds on the level of bias expected from the T08 fits.

\begin{figure}
\includegraphics[width=0.45\textwidth]{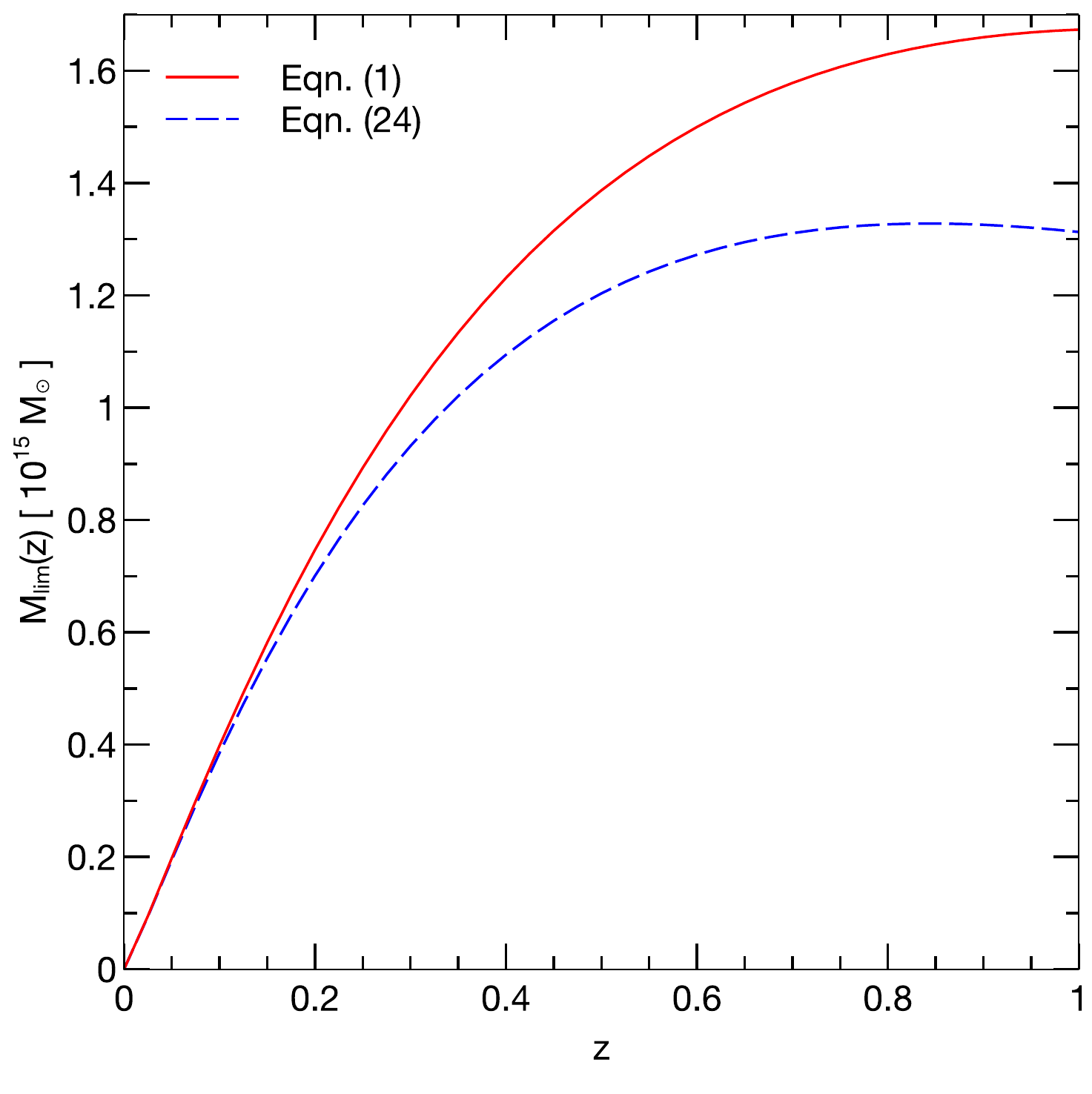}
\caption{The two Planck-like selection thresholds used in the text, \eqn{Moblim-Planck} shown by the solid red curve and \eqn{Moblim-PlanckII} shown by the dashed blue curve. The axes are chosen for ease of comparison with Figure~3 of \cite{planck13-XX-SZcosmo}; the dashed blue curve in particular is in quite good agreement with the green curve for the `shallow zone' in Figure~3 of \cite{planck13-XX-SZcosmo}.}
\label{fig:Moblimcompare}
\end{figure}

Finally, as mentioned earlier, the level of bias in the estimate of $\sig_8$ could in principle be sensitive to the detailed form of the selection threshold (see, e.g., the discussion of the SPT-like Monte Carlo catalog at the end of Section~\ref{sec:montecarlo}). To assess the strength and direction of the effect for the $N$-body halo catalog, we have repeated the $N$-body analysis with the T08 mass function using the following slightly modified form of \eqn{Moblim-Planck}:
\begin{align}
M_{\rm ob,lim}^{\rm (Planck,mod)}(z) &= M_{\rm ob,lim}^{\rm (Planck)}(z)\times(1+z)^{-0.35}\,.
\label{Moblim-PlanckII}
\end{align}
Figure~\ref{fig:Moblimcompare} compares the expressions in \eqns{Moblim-Planck} and~\eqref{Moblim-PlanckII}, with the latter allowing more objects at higher redshifts. We find that the threshold \eqref{Moblim-PlanckII} with its \emph{ad hoc} factor $(1+z)^{-0.35}$ is in better agreement than \eqn{Moblim-Planck} with the threshold reported in \cite{planck13-XX-SZcosmo} (compare Figure~\ref{fig:Moblimcompare} with the green curve for the `shallow zone' in Figure~3 of \cite{planck13-XX-SZcosmo}). The threshold \eqref{Moblim-PlanckII} leads to catalogs with $1046$ clusters on average for the $m_{\rm 200b}$ case and $167$ clusters on average for the $m_{\rm 500c}$ case.

%

\subsection{Results}
\label{sec:Nbody:sub:results}
\begin{figure}
\includegraphics[width=0.485\textwidth]{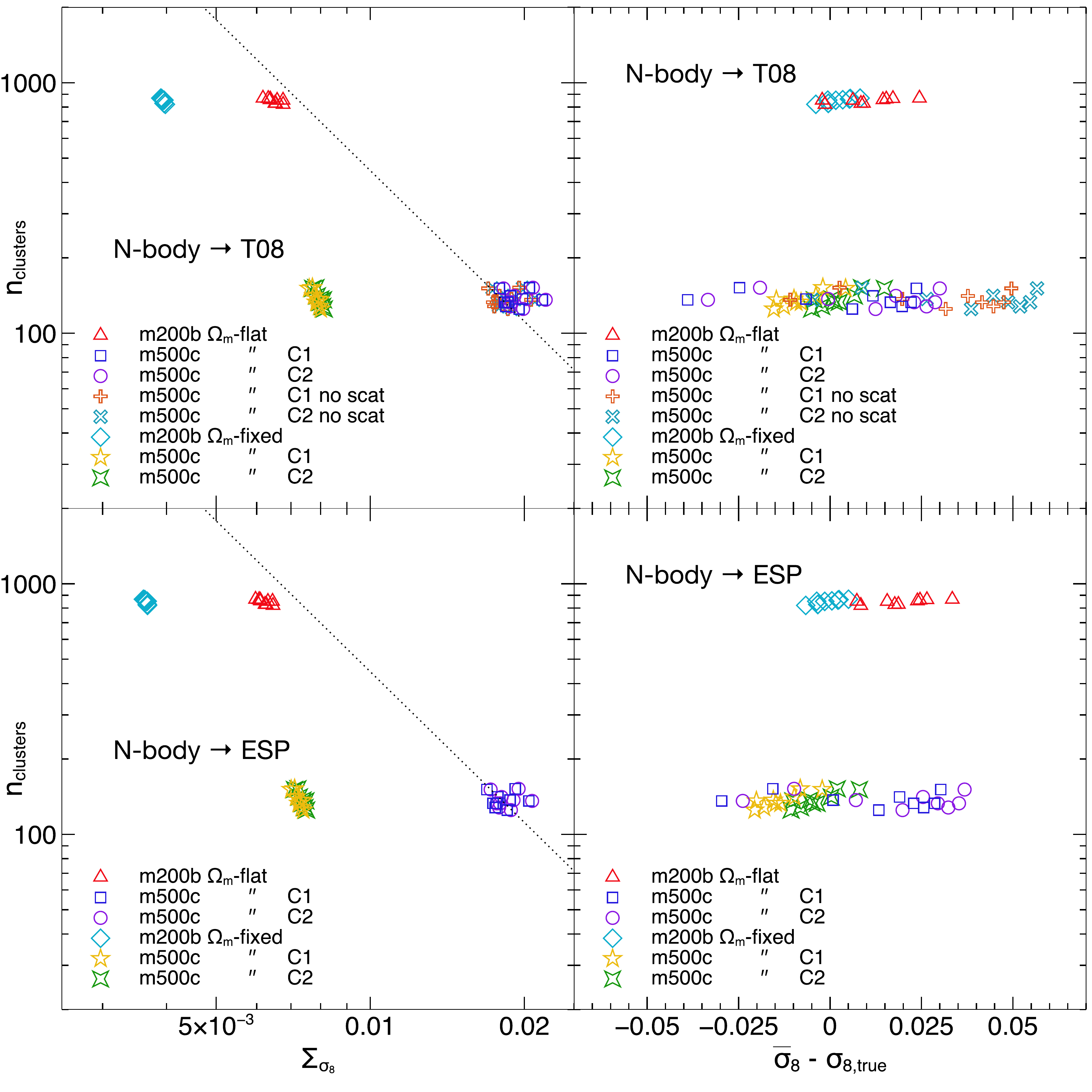}
\caption{Joint distributions of the number of mock clusters $n_{\rm clusters}$ with the standard deviation $\Sigma_{\sig_8}$ of the posterior $p(\sig_8)$ \emph{(left panels)} and with the absolute bias $\bar\sig_8-\sig_{\rm 8,fid}$ \emph{(right panels)} as measured in the $N$-body based mock catalogs when using the T08 \emph{(top panels)} and ESP \emph{(bottom panels)} mass functions for the likelihood analysis. The format of the left and right panels is similar to the corresponding panels of Figure~\ref{fig:ncdist}, and the dotted line in the left panels is the same as that in the left panel of Figure~\ref{fig:ncdist}. Results are shown for the various combinations of mass definition and prior on $\Om_{\rm m}$ as displayed in Table~\ref{tab:Nbody}, for the survey selection threshold \eqref{Moblim-Planck}. Note in particular that the values of absolute bias $\bar\sig_8-\sig_{\rm 8,fid}$ for the case when the intrinsic scatter of $p(\ln m_{\rm 500c}|\ln m_{\rm 200b})$ is ignored (labelled ``no scat'') are systematically more positive than all other cases (see also Table~\ref{tab:Nbody}).}
\label{fig:ncdistNbody}
\end{figure}

\noindent
Since we have only $9$ $N$-body realisations, instead of showing cumulative histograms we simply quote mean values of $s$ with associated errors for various cases in Table~\ref{tab:Nbody}.
Figure~\ref{fig:ncdistNbody} has a similar format as Figure~\ref{fig:ncdist} and shows joint distributions of $n_{\rm clusters}$, $\Sigma_{\sig_8}$ and $\bar\sig_8-\sig_{8,{\rm fid}}$ for the $9$ $N$-body catalogs when analysed using the T08 (top row) and ESP (bottom row) mass functions. Note that the individual points in each cloud in the top row can be compared with the corresponding points in the bottom row, since we analysed the same clusters using T08 and ESP in each realisation. 

The last row in Table~\ref{tab:Nbody} shows the results for $s$ when using \eqn{Moblim-PlanckII}. We see that the numbers are all systematically higher than those using \eqn{Moblim-Planck}; this is consistent with the fact that the typical absolute bias $\bar\sig_8-\sig_{8,{\rm fid}}$ changes relatively little compared to the decrease in the typical standard deviation $\Sigma_{\sig_8}$ due to having more objects.
Table~\ref{tab:Nbody-absbias} gives the mean values of the absolute bias $100\times\avg{(\bar\sig_8 - \sig_{8,{\rm fid}})}$ for all the cases that appear in Table~\ref{tab:Nbody}. Notice that in this case the mean values for the T08 analysis in the first and third rows are nearly identical, showing that our results for the absolute bias are not very sensitive to changes in the selection threshold; only with dramatic differences such as those between the Planck and SPT selection functions would we expect significant changes in the absolute bias (c.f. Section~\ref{sec:montecarlo}).

Table~\ref{tab:Nbody} shows that, for any combination of mass function, mass definition and calibration scheme, the mean significance $\avg{s}$ shifts to more negative values as we tighten the prior on $\Om_{\rm m}$. This is identical to the trend seen when analysing ESP mock clusters using T08 for the $m_{\rm 500c}$ case.
The relative trends between T08 and ESP are also mostly similar to those seen in the Monte Carlo mocks in the previous Section, with some differences. 
E.g., as for the mocks, the T08 mass function with $m_{\rm ob}=m_{\rm 200b}$  leads to a more positive mean significance $\avg{s}$ than ESP when $\Om_{\rm m}$ is fixed to its true value, while this is reversed when using a flat prior on $\Om_{\rm m}$. Similarly, for the $m_{\rm 500c}$ case, both calibration schemes $C_1$ and $C_2$ lead to lower values of $\avg{s}$ for T08 than for ESP when using a flat $\Om_{\rm m}$ prior. 
For the $m_{\rm 500c}$ case with fixed $\Om_{\rm m}$, however, the relative trend is opposite to the one seen in the mocks: $\avg{s}$ for T08 is larger than for ESP for both calibration schemes. (In fact this is the same as for the corresponding case with $m_{\rm 200b}$.) 

It is clear, however, that the T08 mass function does not under-estimate the value of $\sig_8$ (which would have to be the case in order to explain the current tension in the Planck data). 
For the $m_{\rm 500c}$ analysis with a flat $\Om_{\rm m}$ prior, if we ignore the intrinsic scatter of $p(\ln m_{\rm 500c}|\ln m_{\rm 200b})$, we see that the T08 mass function \emph{over-estimates} $\sig_8$ by as much as $2$-$2.5\sig$ depending on the choice of calibration scheme and selection threshold. These numbers should be upper limits for the bias in real surveys as discussed above. Accounting for the scatter leads to results that are consistent with being unbiased (for both calibration schemes and selection thresholds), although still showing a mild tendency to over-estimate $\sig_8$. And this tendency is enhanced again for the case $m_{\rm ob} = m_{\rm 200b}$. 

As discussed earlier, the primary difference between the $m_{\rm 200b}$ and $m_{\rm 500c}$ cases is the difference in number of clusters $n_{\rm clusters}$ analysed and hence in the typical values of the width $\Sigma_{\sig_8}$ of the posterior distribution of $\sig_8$, while the typical values of absolute bias $\bar\sig_8-\sig_{8,{\rm fid}}$ differ less, although they are more positive for $m_{\rm 200b}$ (c.f. Table~\ref{tab:Nbody-absbias}). This trend of larger bias for $m_{\rm 200b}$ is seen for both choices of prior on $\Om_{\rm m}$. 
This points to the need for calibrating a new mass function estimate, tailored for cluster cosmology, that would remain unbiased even for the larger $n_{\rm clusters}$ values expected from upcoming data releases. We discuss this further in Section~\ref{sec:conclude} below.

The ESP mass function, while faring systematically somewhat worse than T08, nevertheless leads to comparable biases. 
For the flat $\Om_{\rm m}$ prior, for each choice of mass definition and calibration, the value of $\avg{s}$ with ESP is roughly a factor $2$ times the corresponding value with T08. For the fixed $\Om_{\rm m}$ case, $\avg{s}$ with ESP is shifted compared to T08 by $\sim-0.8$.

\begin{table*}
\centering
\begin{tabular}{|c|c|c|cc|cc|c|cc|}
\hline
& & \multicolumn{5}{c|}{T08} & \multicolumn{3}{c|}{ESP} \\
\hline
& & $m_{\rm 200b}$ & \multicolumn{2}{c|}{$m_{\rm 500c}$} & \multicolumn{2}{c|}{$m_{\rm 500c}$ {\scriptsize (no scatter)}} & $m_{\rm 200b}$ & \multicolumn{2}{c|}{$m_{\rm 500c}$}\\
\hline
selection & $\Om_{\rm m}$ prior &  & $C_1$ & $C_2$ & $C_1$ & $C_2$ & & $C_1$ & $C_2$ \\
\hline\hline
Eqn.~\eqref{Moblim-Planck} & flat   & $+1.0\pm0.3$ & $+0.3\pm0.7$ & $+1.0\pm0.7$ & $+2.9\pm0.7$ & $+3.6\pm0.7$ & $+2.0\pm0.3$ & $+1.1\pm0.7$ & $+1.7\pm0.7$ \\[1pt]
& fixed  & $+0.2\pm0.1$ & $-0.8\pm0.2$ & $+0.2\pm0.2$ & -- & -- & $-0.1\pm0.1$ & $-1.4\pm0.2$ & $-0.4\pm0.2$ \\[1pt]
\hline
Eqn.~\eqref{Moblim-PlanckII} & flat   & $+1.0\pm0.3$ & $+0.5\pm0.5$ & $+1.1\pm0.5$ & $+3.0\pm0.5$ & $+3.7\pm0.5$ & -- & -- & -- \\[1pt]
\hline
\end{tabular}
\caption{Summary of mean values of the absolute bias $100\times\avg{(\bar\sig_8-\sig_{8,{\rm fid}})}$ for the $N$-body based mock catalogs discussed in Section~\ref{sec:Nbody}, with errors given by the standard error over $9$ independent realisations. The format is identical to that of Table~\ref{tab:Nbody}. Notice that the mean values for the T08 analysis in the first and third rows are nearly identical, showing that our results for the absolute bias are not very sensitive to changes in the selection threshold.}
\label{tab:Nbody-absbias}
\end{table*}

\section{Summary and Conclusions}   
\label{sec:conclude}
\noindent
The quality and quantity of cosmological data are now at the stage where systematic effects at the few per cent level can potentially be mistaken for new physics \cite{planck13-XX-SZcosmo,sfh13,hh13,bm13}. In this paper we focused on cosmological analyses using galaxy clusters; these involve several ingredients amongst which the assumed halo mass function plays a key role. We have presented an in-depth statistical analysis to test the performance of two analytical mass function prescriptions, the \citet[][T08]{Tinker08} fit to $N$-body simulations and the Excursion Set Peaks (ESP) theoretical model of \citet{psd13}. Such an analysis is particularly timely in light of recent results showing a $2$-$3\sig$ tension between the values of $\sig_8$ recovered from cluster analyses such as those using the Planck SZ catalog \cite{planck13-XX-SZcosmo} or data from the SPT \cite{reichardt+13}, and the Planck CMB analysis \cite{planck13-XVI-cosmoparams}.

Our basic strategy involved generating mock cluster catalogs and running them through a likelihood analysis pipeline that mimics what is typically used for real data. This includes the conversion between the observable and the true halo mass, which we modelled using two mass definitions $m_{\rm 200b}$ and $m_{\rm 500c}$, treating one as the true mass and the other as the observable and accounting for the relative scatter and mean offset between the two. 

We first used Monte Carlo catalogs generated assuming the ESP mass function to be the `truth', which we analysed using the T08 mass function. This allowed us to explore statistical differences between these mass functions for various choices of observable-mass relations, survey selection criteria and priors on parameters degenerate with $\sig_8$. For example, we showed that although these mass functions agree at the $\sim10\%$ level at any given redshift, for a Planck-like survey the constraints on $\sig_8$ recovered from each could be different by as much as $2\sig$ (see Section~\ref{sec:mc:sub:ESPvsTinker} and Table~\ref{tab:mocks} for details). While we used survey selection thresholds (limiting masses) inspired by Sunyaev-Zel'dovich surveys such as Planck and SPT, our results are also relevant for other surveys with similar limiting masses as a function of redshift.

We then repeated the analysis with mock Planck-like cluster catalogs built using halos identified in $N$-body simulations and organised into lightcones. This is an important consistency check for the T08 mass function fit which is routinely used in galaxy cluster analyses \emph{without} accounting for the errors inherent in the fit parameter values, which could have significant effects due to scatter across the mass selection threshold. Indeed, we saw that ignoring the intrinsic scatter between $m_{\rm 500c}$ and $m_{\rm 200b}$ -- which is similar to (but likely more extreme than) ignoring the scatter due to parameter errors -- leads to an \emph{over-estimation} of the value of $\sig_8$ by as much as $2\sig$ (see the columns marked ``T08 $m_{\rm 500c}$ (no scatter)'' in Table~\ref{tab:Nbody}, and the discussion towards the end of Section~\ref{sec:Nbody:sub:analysis}).  

When the intrinsic scatter is accounted for, the significance of this bias is considerably reduced and the T08 analysis becomes essentially unbiased. 
However, we saw that increasing the number of clusters analysed (by switching from $m_{\rm 500c}$ to $m_{\rm 200b}$ while using the same selection threshold) leads to similar values of the absolute bias $\bar\sig_8-\sig_{8,{\rm fid}}$ while obviously decreasing the typical width $\Sigma_{\sig_8}$ of the $\sig_8$ posterior, thereby leading once again to a systematic over-estimation of $\sig_8$.
Moreover, this $m_{\rm 200b}$ analysis gives a much cleaner comparison between the simulations and the T08 fit, since it avoids making any of the assumptions regarding mass conversion discussed in Appendix~\ref{app:masscal}. Similar trends for the absolute bias and significance are obtained when the selection threshold is altered to allow more objects at higher redshifts (see Tables~\ref{tab:Nbody} and~\ref{tab:Nbody-absbias}, and the discussion in Sections~\ref{sec:Nbody:sub:analysis} and~\ref{sec:Nbody:sub:results}).

We concluded that (a) the T08 fit -- \emph{provided one accounts for the scatter when converting from $m_{\Delta{\rm b}}$ to $m_{\rm 500c}$} -- should be close to unbiased in $\sig_8$ for a current Planck-like survey and (b) with an increased number $n_{\rm clusters}$ as might be expected from upcoming Planck data releases, the T08 fit could lead to $\sig_8$ values biased high at $>1.5\sig$, which would exacerbate the current tension between cluster analyses and the Planck CMB results \cite{planck13-XVI-cosmoparams}.

Additionally, we analysed the $N$-body based catalogs with the ESP mass function, and found that it leads to comparable but systematically more biased results than the T08 fit.
The ESP model, however, was a proof-of-concept example presented by \citet{psd13} with minor tuning and was not intended for high-performance precision cosmology. As we discussed, one of the strongest features in this model is the natural prediction of mild non-universality in the mass function with no free parameters. The T08 fit on the contrary needed several parameters specifically to describe this behaviour of the mass function, since the basic template for this fit was the \emph{universal} prediction of the original excursion set approach. 

In light of our findings above, this suggests that it might now be more economical to build new fitting functions based on the non-universal ESP prescription instead, with the aim of obtaining an analytical function that remains unbiased even in the face of the better quality data that will soon be available. Conceivably, such a fit could be tailored for the high mass regime relevant for specific cluster surveys. 
We leave such a calibration to future work.

\acknowledgments
\noindent
Warm thanks to Oliver Hahn for his technical support with the $N$-body simulations and for many useful conversations and comments on the draft. I gratefully acknowledge the use of computing facilities at ETH Z\"urich, and thank V. Springel, O. Hahn and P. Behroozi for making their respective codes publicly available. I also thank J. Dietrich, S. Seehars and E. Sefusatti for helpful discussions and comments on the draft, A. Ludlow for useful correspondence, and an anonymous referee for comments that helped improve the quality of the paper.

\bibliography{masterRef,refSig8Bias}

\appendix
\section{Mass calibration}
\label{app:masses}
\noindent
In this Appendix we discuss various issues related to mass calibration. We start in \ref{app:masscal} by describing how we calibrate the distribution $p(\ln m_{\rm 500c}|\ln m_{\rm 200b})$, accounting for its cosmology dependence. Then in \ref{app:restest} we perform a resolution study for the halo identification step and assess the resulting effects on $p(\ln m_{\rm 500c}|\ln m_{\rm 200b})$.

\begin{figure*}
\includegraphics[width=0.425\textwidth]{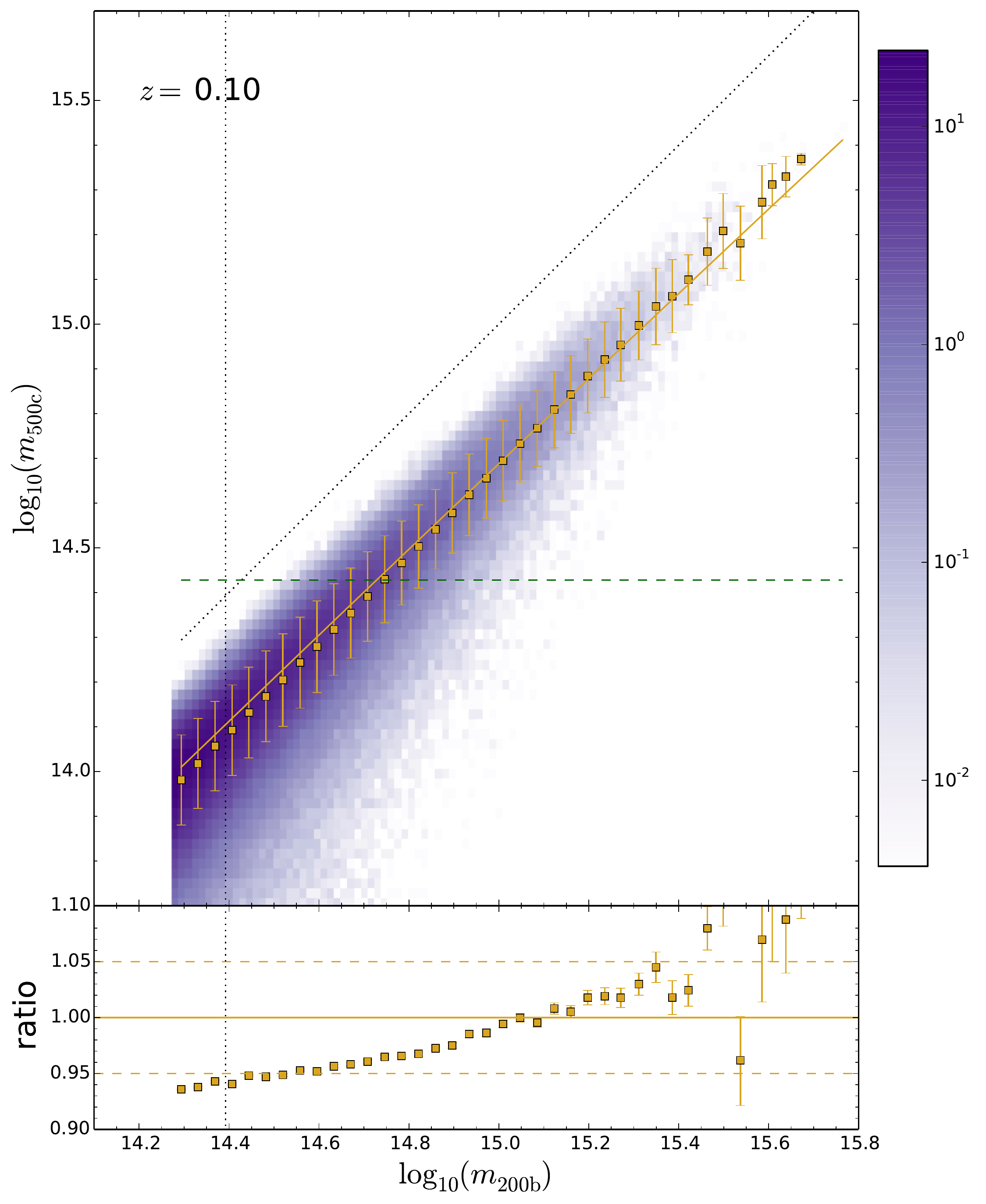}
\hskip 0.05\textwidth
\includegraphics[width=0.425\textwidth]{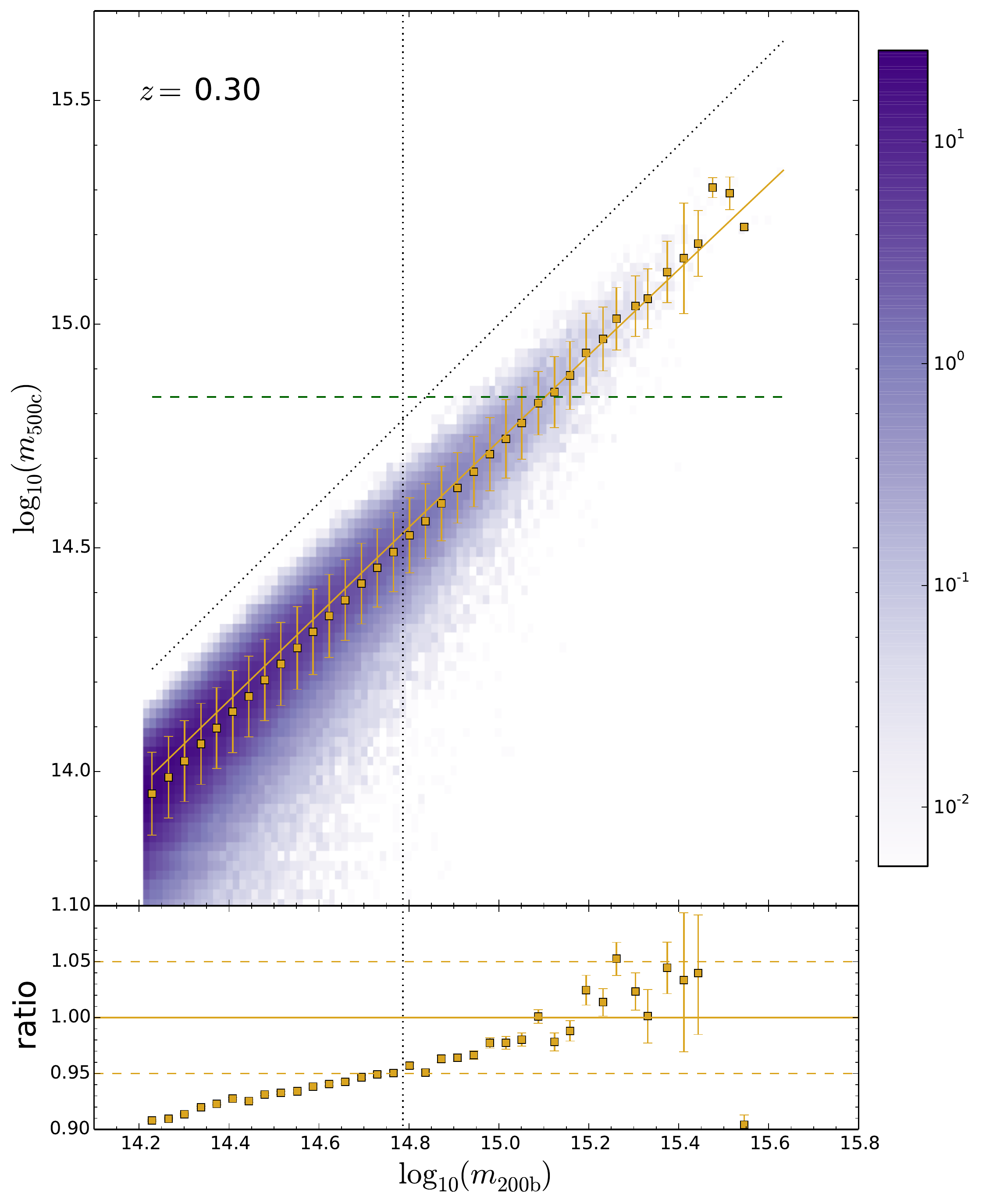}
\caption{Mass calibration using scheme $C_1$. \emph{(Top panels):} Coloured region shows the normalised joint distribution of $\log m_{\rm 500c}$ and $\log m_{\rm 200b}$ in our $9$ $N$-body realisations at redshift $z=0.1$ \emph{(left panel)} and $z=0.3$ \emph{(right panel)}. The yellow squares show the median of $\log m_{\rm 500c}$ in bins of $\log m_{\rm 200b}$ (with the bin ``center'' being defined as the median value of $\log m_{\rm 200b}$ in the bin) and error bars show the standard deviation of $\log m_{\rm 500c}$ in the bin. The solid yellow line shows the analytical calculation for the conditional expectation value $\avg{\log m_{\rm 500c}|\log m_{\rm 200b}}$ using the procedure described in the text, where we set the mean concentration-mass-redshift relation to be \eqn{app-meanc200b} with $\alpha=9.0,\beta=0.4$, which we call scheme $C_1$. The diagonal dotted black line shows the one-to-one relation; the mean relation is clearly significantly biased. The horizontal dashed green line shows the limiting value of $m_{\rm 500c}$ at the corresponding redshift as given by the Planck-like selection threshold~\eqref{Moblim-Planck}. The vertical dotted black like indicates the value of $m_{\rm 200b}$ at which the horizontal line lies $3\sig_{\log m_{\rm 500c}}$ away from the measured median $\log m_{\rm 500c}$ of the bin, giving a rough indication of the range of $m=m_{\rm 200b}$ values that contribute to the integral over $m$ in \eqn{mu-mobz}.
\emph{(Bottom panels):} Ratio of the measured value of $10^{\log m_{\rm 500c}|_{\rm median}}$ to the analytical calculation $10^{\avg{\log m_{\rm 500c}|\log m_{\rm 200b}}}$. The error bars in this case are the \emph{standard error} computed over all halos that contribute to the bin, and are therefore typically significantly smaller than the corresponding scatter shown in the top panels. Clearly, the scheme $C_1$ is accurate at $\lesssim5\%$ over the mass range of interest; similar results are true at other redshifts as well.}
\label{fig:masscalC1}
\end{figure*}

\subsection{Calibrating $p(\ln m_{\rm 500c}|\ln m_{\rm 200b})$}
\label{app:masscal}
\noindent
We will assume throughout that CDM halos follow the NFW \cite{nfw96} density profile specified for a chosen mass definition, say $m_{\Delta{\rm b}}$, and the corresponding concentration parameter $c_{\Delta{\rm b}}$ as
\begin{align}
&\rho(r|m_{\Delta{\rm b}},c_{\Delta{\rm b}}) = \frac{\rho_s}{(r/r_s)(1+r/r_s)^2}\,,\notag\\
&m_{\Delta{\rm b}} = 4\pi\rho_sR_{\Delta{\rm b}}^3f(1/c_{\Delta{\rm b}}) = \frac{4\pi}{3}R_{\Delta{\rm b}}^3\Delta\bar\rho\,,\notag\\
&c_{\Delta{\rm b}} = \frac{R_{\Delta{\rm b}}}{r_s}\,;\,\,f(x) \equiv x^3\left(\ln(1+x^{-1})-(1+x)^{-1}\right)\,,
\label{app-nfwprofile}
\end{align}
with analogous definitions for $m_{\Delta{\rm c}}$ and $c_{\Delta{\rm c}}$, where one must also replace $\bar\rho$ with $\rho_{\rm c}$ in the second line. (The function $f$ should not be confused with the mass fraction discussed in the main text.) 
For fixed values of mass and concentration, \citet{hk03} gave an accurate analytical prescription to convert from one mass definition $m_{\Delta_1}$ to another $m_{\Delta_2}$, where $\Delta_1$ and $\Delta_2$ can each be defined with respect to either the mean density or critical density. We will use this prescription below.

\begin{figure*}
\includegraphics[width=0.425\textwidth]{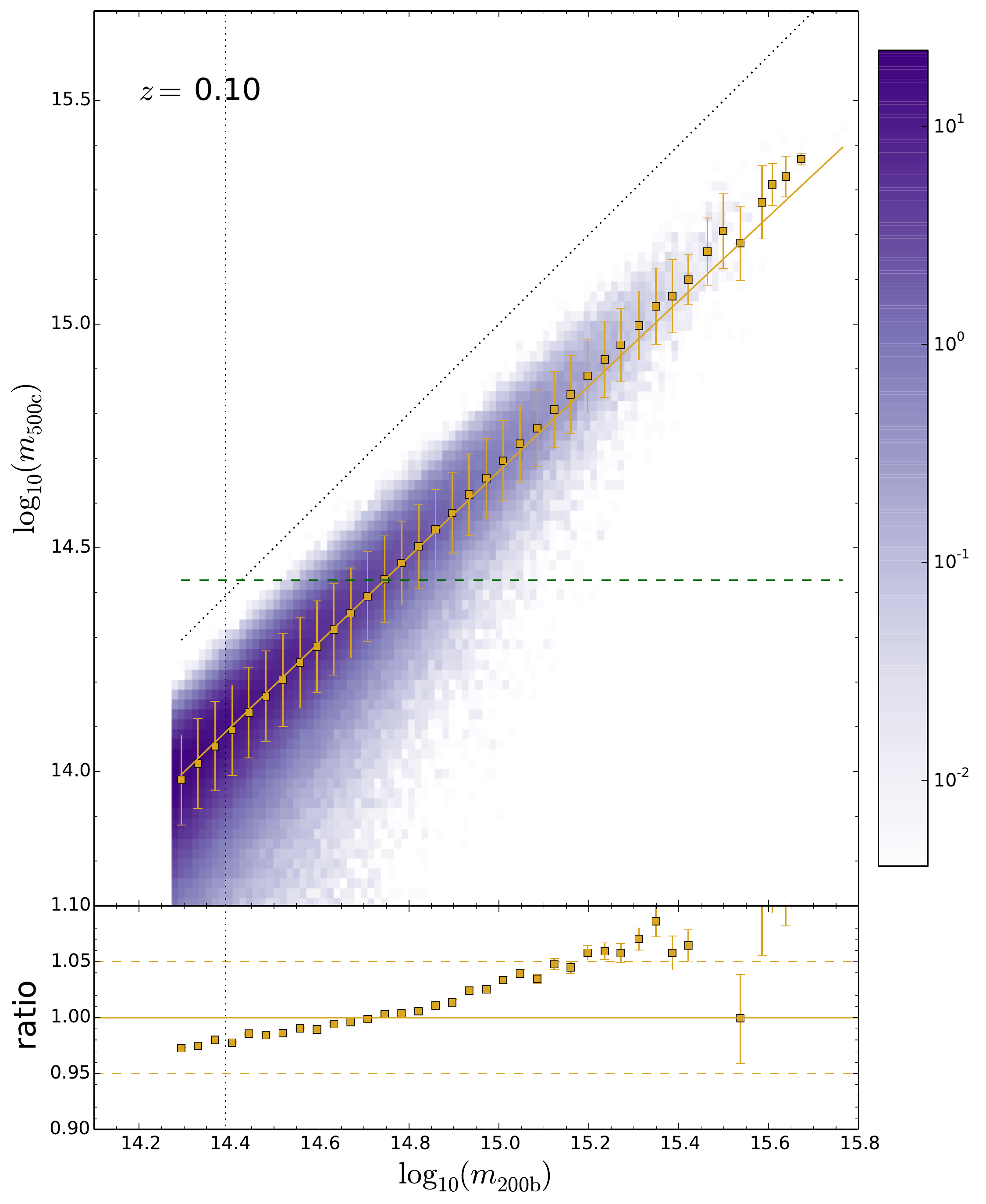}
\hskip 0.05\textwidth
\includegraphics[width=0.425\textwidth]{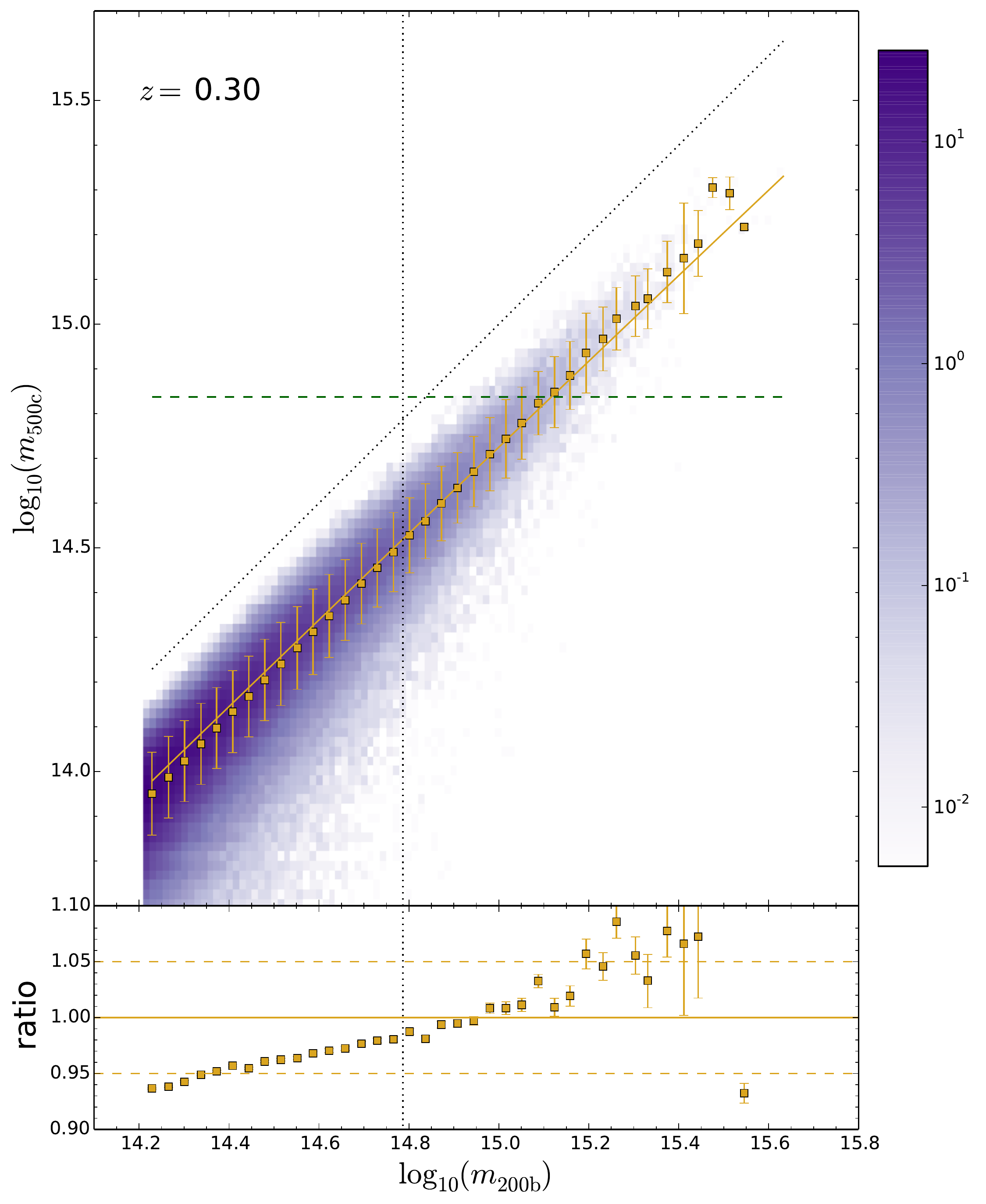}
\caption{Mass calibration using scheme $C_2$. Comparing the same data as in Figure~\ref{fig:masscalC1} with the analytical calculation using the calibration scheme $C_2$ where we set $\alpha=8.0,\beta=0.375$ in \eqn{app-meanc200b}. The format is identical to Figure~\ref{fig:masscalC1} and the results are qualitatively similar, except that the scheme $C_2$ describes the data better at lower masses than $C_1$.}
\label{fig:masscalC2}
\end{figure*}

The concentration parameter is not predicted by theory and must be measured in simulations (although see \cite{ludlow+13,jsdm14-II}). Typical calibrations in the literature exist for $c_{\rm 200c}(m_{\rm 200c},z)$ which shows a Lognormal distribution at fixed mass with a width $\sig_{\ln c}\simeq0.18$ (which we adopt here) that is approximately independent of cosmology, redshift and mass \cite{bullock+01,dolag+04}. For the mean value $\avg{\ln c_{\rm 200c}}$ there are many results \cite{bullock+01,dolag+04,duffy+08,gao+08,prada+12,giocoli+12,ludlow+12,kwan+13,ad13}, a particularly interesting recent result being that of \citet{ludlow+13} which is reasonably well described (within $\sim10\%$ for the mass and redshift ranges relevant for us) by $\bar c_{\rm 200c}(m_{\rm 200c},z) \equiv {\rm e}^{\avg{\ln c_{\rm 200c}}} = \alpha\, \nu(m_{\rm 200c},z)^{-\beta}$ for relaxed halos, where $\nu(m,z)$ was defined in \eqn{nu-def} and where $\alpha\simeq6$ and $\beta\simeq0.4$ nearly independent of cosmology. Here we wish to calibrate the distribution $p(\ln m_{\rm 500c}|\ln m_{\rm 200b})$, so it will be more useful to have results for $c_{\rm 200b}(m_{\rm 200b},z)$. In principle we should measure $c_{\rm 200b}$ for our halos and thereby determine its mean value, but for simplicity we choose a more phenomenological approach. Motivated by the results of \cite{ludlow+13} we assume the form 
\be
\bar c_{\rm 200b}(m_{\rm 200b},z) \equiv {\rm e}^{\avg{\ln c_{\rm 200b}}} = \alpha\, \nu(m_{\rm 200b},z)^{-\beta}\,,
\label{app-meanc200b}
\ee
and choose the values of $\alpha$ and $\beta$ such that the mass calibration achieves (by trial and error) a desired accuracy as described below. In this form, the concentration depends weakly on $\Om_{\rm m}$ and $\sig_8$ which is consistent with the results of \cite{dolag+04,ludlow+13} for the relevant mass and redshift ranges.
We proceed with the caveat, however, that the assumption of the NFW profile may not be very accurate for all the massive halos we are interested in, since these may not all be in fully relaxed dynamical states \cite{gao+08,ludlow+12}.

For a fixed value of $m_{\rm 200b}$ and $c_{\rm 200b}$, the Hu-Kravtsov prescription gives a unique value $m_{\rm 500c}(m_{\rm 200b},c_{\rm 200b})$ through the following relations (c.f. equations C7-C11 of \cite{hk03}): 
\begin{align}
&m_{\rm 500c} = \frac{4\pi}{3}R_{\rm 500c}^3\times500\times\rho_{\rm c}(z)\,,\notag\\ 
&m_{\rm 200b} = \frac{4\pi}{3}R_{\rm 200b}^3\times200\times\Om_{\rm m}(z)\rho_{\rm c}(z)\,,\notag\\
&f(r_s/R_{\rm 500c}) = \frac{500}{200\Om_{\rm m}(z)}\times f(1/c_{\rm 200b}) \equiv f_{\rm 500c}\,,\notag\\
&\frac{r_s}{R_{\rm 500c}} = x\left(f_{\rm 500c}\right)\,\notag\\
&\frac{m_{\rm 500c}}{m_{\rm 200b}} = \frac{500}{200\Om_{\rm m}(z)}\left(\frac{R_{\rm 500c}}{c_{\rm 200b}r_s}\right)^3\,.
\label{app-massconversion}
\end{align}
\citet{hk03} showed that the inverse function $x(f)$ needed in the fourth line is accurately approximated by
\be
x(f) = 2f + \left[a_1f^{2p}+0.5625\right]^{-1/2}
\label{app-x-of-f}
\ee
where $p=a_2+a_3\ln f+a_4(\ln f)^2$ with $(a_1,a_2,a_3,a_4)=(0.5116,-0.4283,-3.13\times10^{-3},-3.52\times10^{-5})$.

Since the concentration is a stochastic quantity, however, we need an additional step to account for the scatter in $c_{\rm 200b}$ at fixed mass $m_{\rm 200b}$. We do this by numerically marginalising the value of $\ln m_{\rm 500c}$ as obtained above, over a Gaussian distribution in $\ln c_{\rm 200b}$ with mean $\ln \bar c_{\rm 200b}$ (equation~\ref{app-meanc200b}) and standard deviation $\sig_{\ln c}=0.18$ as described above. This gives us the expectation value  $\avg{\ln m_{\rm 500c}|\ln m_{\rm 200b}}$ of the distribution $p(\ln m_{\rm 500c}|\ln m_{\rm 200b})$, which explicitly depends on $\Om_{\rm m}$ as can be seen from \eqn{app-massconversion}.

The prescription above should correctly account for the cosmology dependence of the mass conversion on average. The scatter in the mass conversion is expected to be approximately independent of cosmology, since this is closely linked to the scatter in the concentration which is seen to be nearly constant as mentioned above. We therefore account for the width of $p(\ln m_{\rm 500c}|\ln m_{\rm 200b})$ in a more approximate way as follows.
We assume that $p(\ln m_{\rm 500c}|\ln m_{\rm 200b})$ is Gaussian in $\ln m_{\rm 500c}$ (i.e., Lognormal in $m_{\rm 500c}$). This is mainly done for simplicity, although in the relevant mass range this is not a bad approximation. We assume that the standard deviation $\sig_{\ln m_{\rm 500c}}$ of this distribution is independent of cosmology, and use the value measured in our simulations; this varies slowly with redshift with $\sig_{\ln m_{\rm 500c}}(z=0.1)=0.20$ and $\sig_{\ln m_{\rm 500c}}(z=1.0)=0.16$ (the functional form $\sig_{\ln m_{\rm 500c}}(z)=0.20\times(1+z)^{-0.33}$ decribes our measurements to within $5\%$ over this redshift range). We additionally choose to ignore the redshift dependence of $\sig_{\ln m_{\rm 500c}}$ as well, and simply use the mean value of $(\sig_{\ln m_{\rm 500c}}(z))^2$ over our $10$ snapshots (which is $(0.179)^2$) as the variance of $p(\ln m_{\rm 500c}|\ln m_{\rm 200b})$. Finally, for the Planck-like survey we include an additional scatter in quadrature of $10\%$ in mass, as discussed in the text, to get $\sig_{\ln m_{\rm ob}}=0.21$, while for the SPT-like survey we add a $20\%$ scatter to get $\sig_{\ln m_{\rm ob}}=0.27$.

The calibration of the mean value $\avg{\ln m_{\rm 500c}|\ln m_{\rm 200b}}$ depends on the choice of the mean concentration-mass-redshift relation \eqref{app-meanc200b}. To assess the sensitivity of our results to this choice, we perform the analysis for two sets of parameters $\{\alpha,\beta\}$: we label these as the calibration schemes $C_1$ 
and $C_2$ with parameter values given by
\begin{align}
C_1\,&:\quad \alpha=9.0\,;\,\beta=0.4\,,\notag\\
C_2\,&:\quad \alpha=8.0\,;\,\beta=0.375\,.
\label{app-calschemes}
\end{align}
Note that the normalisations $\alpha$ are not the same as for the results of \cite{ludlow+13} for $\bar c_{\rm 200c}$ mentioned above; this is expected since the value of concentration depends on the chosen mass definition. (The normalisations are close to what is predicted by the Hu-Kravtsov prescription applied to the $m_{\rm 200b}$-$m_{\rm 200c}$ conversion, as they should be.) 

Figure~\ref{fig:masscalC1} shows the comparison of the mean value $\avg{\log m_{\rm 500c}|\log m_{\rm 200b}}$ calculated as described above using the scheme $C_1$, with the measured median value of $\log m_{\rm 500c}$ in bins of fixed $\log m_{\rm 200b}$ at $z=0.1,0.3$ using all our simulations (we find qualitatively similar results for all our other snapshots). We also display the measured joint distributions of $m_{\rm 500c}$ and $m_{\rm 200b}$ as the coloured regions, and the limiting mass \eqref{Moblim-Planck} for $m_{\rm 500c}$ as the horizontal green dashed line at each redshift. Figure~\ref{fig:masscalC2} compares the same data with the mean value computed using scheme $C_2$. We see that both schemes $C_1$ and $C_2$ describe the measured median value to within $5\%$ at the relevant masses, with $C_1$ doing better at higher masses and $C_2$ at lower masses.

\subsection{Resolution study}
\label{app:restest}
\noindent
A potential cause for concern is that the smallest relevant masses in our simulations (which are the main drivers of the parameter constraints as discussed in the text) are resolved with about $\sim 400$ particles.
To test whether the values of halo mass returned by {\sc Rockstar} in this case are accurate to a level better than the differences in the calibration schemes discussed above (i.e., better than about $5\%$), we performed a resolution study as described here. 

\begin{figure}
\includegraphics[width=0.485\textwidth]{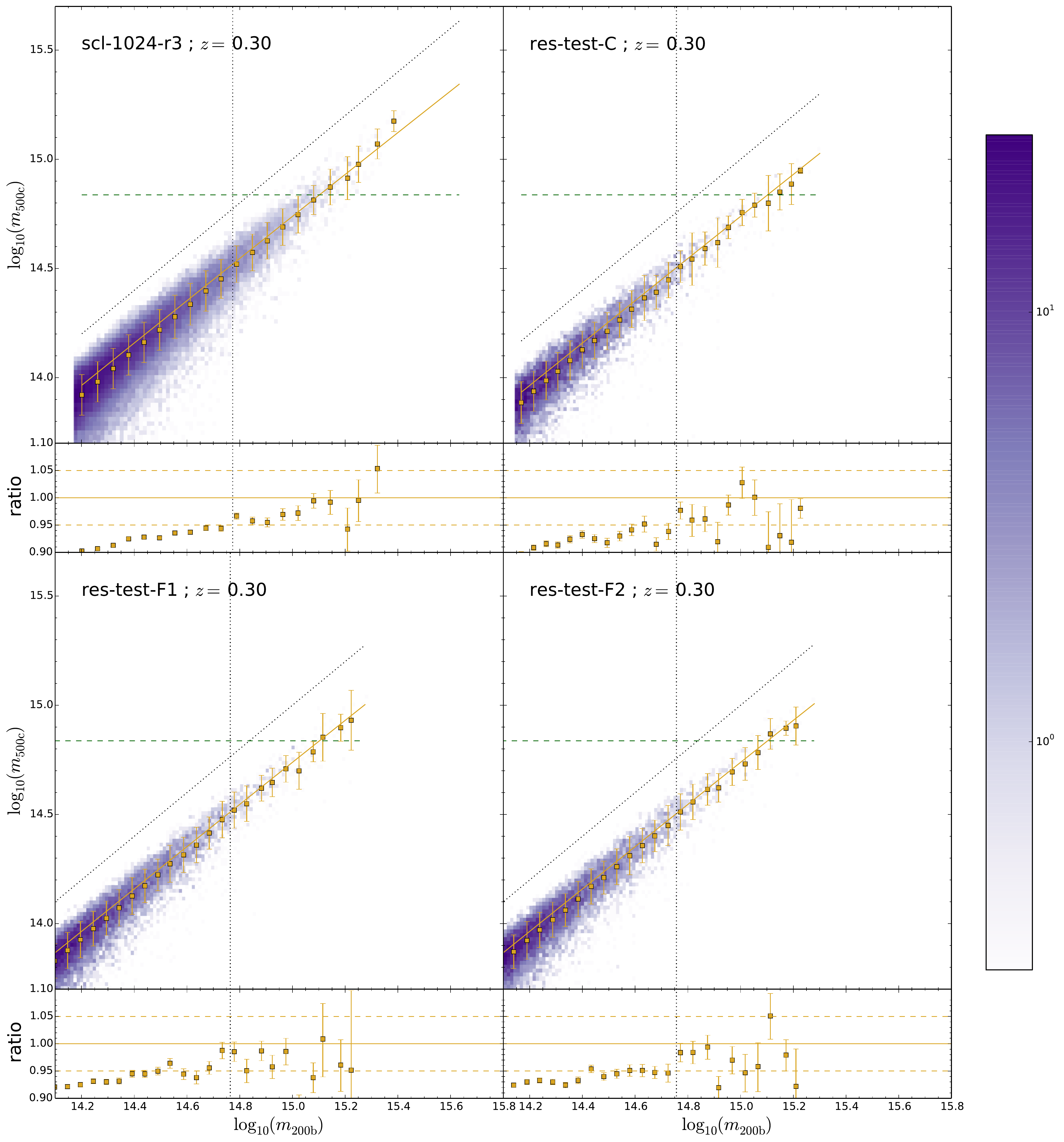}
\caption{Resolution study. The four panels correspond to the following settings for box size, particle resolution and initial conditions generated using \textsc{Music}. \emph{(Top left):} One of the primary runs ($r3$) with $L_{\rm box}=2h^{-1}$Gpc, $N_{\rm part}=1024^3$ and ICs set at resolution level $1024^3$. \emph{(Top right):} `res-test-C' with $L_{\rm box}=1h^{-1}$Gpc, $N_{\rm part}=512^3$ and ICs set at level $512^3$. \emph{(Bottom left):} `res-test-${\rm F}_1$' with $L_{\rm box}=1h^{-1}$Gpc, $N_{\rm part}=1024^3$ and ICs set at two levels, $512^3$ (with the same seed as res-test-C) and $1024^3$. \emph{(Bottom right):} `res-test-${\rm F}_2$' with $L_{\rm box}=1h^{-1}$Gpc, $N_{\rm part}=1024^3$ and ICs set at two levels, $512^3$ (with the same seed as res-test-C) and $1024^3$ (with seed different from the corresponding number in res-test-${\rm F}_1$).
The format of the panels is identical to Figures~\ref{fig:masscalC1} and~\ref{fig:masscalC2}. The solid yellow line in each main panel is the same, corresponding to the analytical value of $\avg{\log m_{\rm 500c}|\log m_{\rm 200b}}$ computed using scheme $C_1$, and is used for the ratio comparison in each sub-panel. The results are shown at redshift $z=0.3$, with qualitatively similar results holding at all other redshifts.}
\label{fig:restest}
\end{figure}

We take advantage of the initial conditions generator {\sc Music} which allows random number seeds to be specified at multiple resolution levels, with the lower resolution random numbers acting as constraints for small scale noise generated at higher resolution. For the primary simulations described in Section~\ref{sec:Nbody:sub:sims} we specify these seeds at a single level ($1024^3$) for each realisation. For the resolution study, we focus on one of these realisations (the one we call $r3$) for which the scatter plot of $m_{\rm 500c}$ against $m_{\rm 200b}$ is shown in the top left panel of  Figure~\ref{fig:restest} (each panel in this Figure has the same format as Figures~\ref{fig:masscalC1} and~\ref{fig:masscalC2}). We display results for a single snapshot at $z=0.3$, but we have checked that our conclusions are true for all $10$ snapshots. The cosmological parameters are fixed at their fiducial values, and the analytical mean value $\avg{\log m_{\rm 500c}|\log m_{\rm 200b}}$ is computed using calibration scheme $C_1$.

Due to computational limitations, we chose to improve mass resolution by using a box size smaller than $L_{\rm box}=2h^{-1}$Gpc. This, however, can potentially introduce volume effects which might be confused with mass resolution effects. To test for this, we first ran an additional box of side $1h^{-1}$Gpc with $512^3$ particles, i.e., using the same mass resolution as the primary simulations, with random number seed specified at the single level of $512^3$. This configuration (denoted `res-test-C') therefore differs from the primary simulation in the box size and dynamic range in mass. However, the smallest relevant halos masses are still resolved with $\sim 400$ particles. We identify halos in this smaller box in the same way as for the primary simulation $r3$ and compare the scatter plot of $m_{\rm 500c}$ against $m_{\rm 200b}$ with the analytical expression for the mean value in the top right panel of Figure~\ref{fig:restest}. The distribution is noisier since there are fewer objects in the smaller volume, but is otherwise consistent with that in the top left panel. In other words, we see no significant volume effects.

We then ran a box of size $1h^{-1}$Gpc with $1024^3$ particles, which therefore has a mass resolution higher than the previous two cases by a factor $8$. To ensure a better comparison, the random number seeds were set at two levels in this case: at the coarse level $512^3$ we used the same seed as in res-test-C, while at $1024^3$ we used the same seed as in the primary run $r3$ (although the latter could also be chosen arbitrarily). This configuration (denoted `res-test-${\rm F}_1$') therefore tests the effects of mass resolution, since it has the same large scale modes and is subject to the same box size effects as res-test-C, but has a better sampling of small scale noise. The mass scatter plot in the bottom left panel of Figure~\ref{fig:restest} shows a minor elevation ($\lesssim2\%$) in the median value of $\log m_{\rm 500c}$ at fixed $m_{\rm 200b}$ as compared to the top panels.

Finally, to assess the effect of \emph{changing} the small scale noise, we ran an additional box with a configuration (denoted `res-test-${\rm F}_2$') which is identical to res-test-${\rm F}_1$ in all respects except for the random number seed at level $1024^3$. The mass scatter plot in the bottom right panel of Figure~\ref{fig:restest} shows results qualitatively very similar to those for res-test-$F_2$.

Overall, then, we see that resolution effects in the masses assigned to halos are smaller than the effect of changing the analytical mass calibration scheme (compare Figure~\ref{fig:restest} with Figures~\ref{fig:masscalC1} and~\ref{fig:masscalC2}).

\section{Generating lightcones}
\label{app:lightcones}
\begin{figure}
\includegraphics[width=0.485\textwidth]{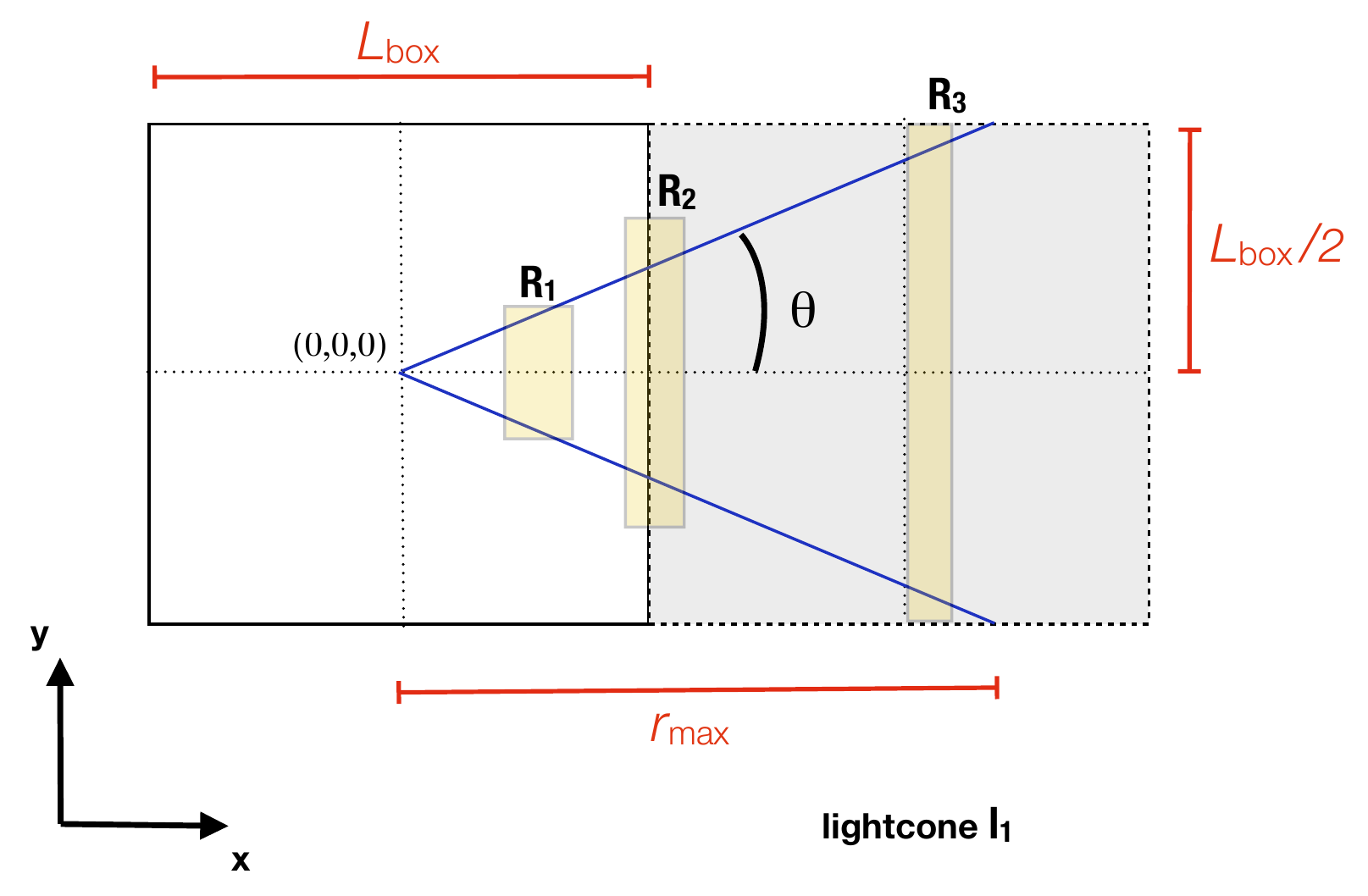}
\caption{An illustration of one of the $6$ cones that make up each lightcone (here, $l_1$). The observer is placed at the center of the box (labelled $(0,0,0)$). The opening angle $\theta$ is chosen so that the last bin of interest at $z=1.0$ is entirely contained in the first periodic copy of the box (shown as the shaded cube) along the axis of the cone, and intersects the adjacent faces in this case at $x=r_{\rm max}=r_{\rm com}(z_{\rm max}+\Delta z/2)$. The yellow shaded rectangles indicate the approximate positions and comoving widths of bins at redshift $z=0.2,0.4,0.9$ from left to right.
To minimise repetition of halos in the full lightcone $l_1$, halos are assigned from realisation $R_1$ when the redshift bin is entirely contained in the main box, from $R_2$ for bins straddling the main box and the first copy, and from $R_3$ for bins entirely in the first copy. See text for more details.}
\label{fig:lightcone}
\end{figure}

\noindent
The halos identified in each snapshot of each realisation were organised into $9$ lightcones as described here. For each box, we place the observer at the center and build six cones around the lines of sight joining the observer perpendicularly to each of the six faces. Figure~\ref{fig:lightcone} shows an example of one of these.

Since our box length is $L_{\rm box}=2h^{-1}$Gpc, the perpendicular comoving distance from the observer to each face corresponds to a redshift $z=z_{\rm box}\simeq0.37$ for this cosmology, beyond which we must account for the periodicity of the simulation. To extend the lightcones to redshift $z=1$, we therefore restrict the opening angle of each of the six cones to $\theta=22.82^\circ$. With this angle, the outermost lines of sight of, e.g., the cone centered around the positive $x$-axis intersect the adjacent faces (e.g., $y=\pm L_{\rm box}/2$, see Figure~\ref{fig:lightcone}) only at $x=r_{\rm max}\equiv r_{\rm com}(z=1.05)=1.19\times L_{\rm box}$ which lies in the first copy of the box along this axis. Here $r_{\rm com}(z) = \int_0^z\der z'\,H(z')^{-1}$ is the comoving distance to redshift $z$ computed for the cosmology of the simulation. 
While this avoids repeating halos that would have contributed to the same cone at redshifts $z < z_{\rm box}$, it still leaves us with a repetition of halos between redshifts $z_{\rm box} < z < 1$ which would contribute, e.g., to the diametrically opposite cone at redshifts $z < z_{\rm box}$, which we get around as follows. 

We take advantage of the fact that we are only interested in $1$-point statistics (the mass function) of the dark matter field, and not in any higher point correlations. We therefore treat multiple realisations of our simulation as being \emph{adjacent} to each other. This would be a very bad approximation if we needed to compute correlation functions straddling the boundary; in the case of the mass function, however, this assumption gives us an explicit realisation of independent redshift bins, which is in fact already assumed in constructing the likelihood \eqref{likelihood}. 

More precisely, we note that there are three distinct types of redshift bins we must deal with for each of the six cones in our chosen geometry: (a) those that are entirely contained inside the main box (all bins with upper edge $z^{+} < z_{\rm box}$), (b) bins straddling the boundary between the main box and the first copy along the chosen axis and (c) bins that are entirely contained in the first copy (the ones with \emph{lower} edge $z^{-} > z_{\rm box}$). Figure~\ref{fig:lightcone} illustrates each of these as the shaded yellow rectangles.

We therefore choose three realisations of the simulation and start selecting halos outwards from the observer in, say, the $+x$-axis cone in each of them. Let us denote the realisations as $R_1,R_2,R_3$ and the cones as $l_1,l_2,l_3$. For bins of type (a), this selection is straightforward, with the lightcone $l_1$ getting halos from realisation $R_1$, and so on (see below for the selection procedure). Upon reaching a bin of type (b), we \emph{cyclically permute} the three realisations, so that $l_1$ now gets its halos from the type (b) bins of $R_2$, $l_2$ from $R_3$ and $l_3$ from $R_1$. We maintain this assignment order until we reach bins of type (c), at which point we cyclically permute once more, so that $l_1$ now gets its type (c) bins from $R_3$, and so on. We repeat the same procedure for all six cones in the box using the same realisations $R_1$-$R_3$. This minimises repetition of any halo in a given lightcone.
Clearly, this procedure can be applied to an arbitrary number of groups of three realisations each, and we perform the analysis for three such groups, thus explaining our choice of $9$ realisations. 

To populate a given redshift bin with central redshift $z_j$, we select halos from the snapshot at $z_j$ of the appropriate realisation such that their centers of mass lie inside the appropriate cone angle at a comoving distance from the center in the range $(r_{\rm com}(z_j-\Del z/2),r_{\rm com}(z_j+\Del z/2))$. Having chosen the halos which satisfy these geometrical constraints, we add a $10\%$ Lognormal scatter to the chosen mass proxy $m_{\rm ob}$, which is one of $m_{\rm 200b}$ or $m_{\rm 500c}$.
Finally, we select halos that satisfy $m_{\rm ob} > M_{\rm ob,lim}^{\rm (Planck)}(z_j)$ for each bin.

We are therefore approximating the mass function as being constant in time across the redshift range spanned by the bin. While this would be a reasonable approximation for small bin widths, in our case $\Del z=0.1$ and this could lead to a significant source of error in the likelihood analysis. To account for this assumed piece-wise constancy of the mass function, we alter the likelihood analysis as described in Section~\ref{sec:Nbody:sub:analysis} (essentially by \emph{modeling} the mass function to be piece-wise constant in exactly the same way). 

The total angle finally spanned by the combination of the $6$ cones in each box amounts to $f_{\rm sky}=0.235$. Our $N$-body sample therefore consists of $9$ mix-matched realisations (which we refer to as our lightcones), each spanning approximately half of the sky area currently analysed by the Planck Collaboration. 

\end{document}